\documentclass[nonacm, sigconf, screen]{acmart}

\renewcommand\footnotetextcopyrightpermission[1]{} 

\makeatletter
\renewcommand\@formatdoi[1]{\ignorespaces}
\makeatother

\AtBeginDocument{%
  }

\usepackage{booktabs}
\usepackage{musicography}
\usepackage{graphicx}
\usepackage{svg}
\usepackage{soul}
\usepackage{enumitem}
\usepackage{wrapfig}
\usepackage{cleveref}
\usepackage{array}
\usepackage{graphicx}
\usepackage{subcaption}
\usepackage{fontawesome5}
\usepackage{xcolor}
\usepackage{colortbl}

\usepackage{xcolor}
\usepackage[normalem]{ulem}
\usepackage{environ}

\newif\ifdraftmode
\draftmodefalse

\definecolor{addedcolor}{HTML}{008080}   
\definecolor{deletedcolor}{HTML}{CCCCCC} 

\newcommand{\added}[2][]{%
  \ifdraftmode
    \leavevmode\textcolor{addedcolor}{\mbox{}#2}%
  \else
    \leavevmode#2%
  \fi
}

\newcommand{\deleted}[2][]{%
  \ifdraftmode
    \leavevmode\textcolor{deletedcolor}{\sout{\mbox{}#2}}%
  \else
  \fi
}

\NewEnviron{addedblock}{%
  \ifdraftmode
    \color{addedcolor}%
    \BODY
  \else
    \BODY
  \fi
}

\NewEnviron{deletedblock}{%
  \ifdraftmode
    \color{deletedcolor}%
    \BODY
  \else
  \fi
}

\begin{document}

\title[LoopLens: Supporting Search as Creation in Loop-Based Music Composition]{LoopLens: Supporting Search as Creation in Loop-Based Music Composition}

\author{Sheng Long}
\orcid{0009-0000-9752-5898}
\affiliation{%
  \institution{Northwestern University }
  \city{Evanston}
  \country{USA}}
\email{shenglong@u.northwestern.edu}

\author{Atsuya Kobayashi}
\orcid{0009-0003-2535-8148}
\affiliation{%
  \institution{Sony Group Corporation}
  \city{Tokyo}
  \country{Japan}}
\email{Atsuya.Kobayashi@sony.com}

\author{Kei Tateno}
\orcid{0009-0000-8249-2659}
\affiliation{%
  \institution{Sony Group Corporation}
  \city{Tokyo}
  \country{Japan}}
\email{Kei.Tateno@sony.com}

\renewcommand{\shortauthors}{Long et al.}

\newcommand{\graphicon}[1]{\includegraphics[height=1.2\fontcharht\font]{#1}}
\newcommand{\convicon}{\includegraphics[height=1.2\fontcharht\font]{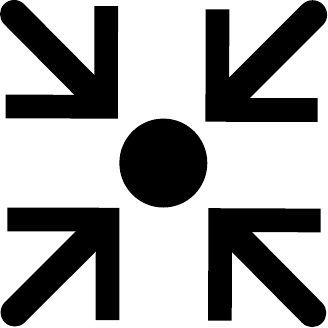}}
\newcommand{\divicon}{\includegraphics[height=1.2\fontcharht\font]{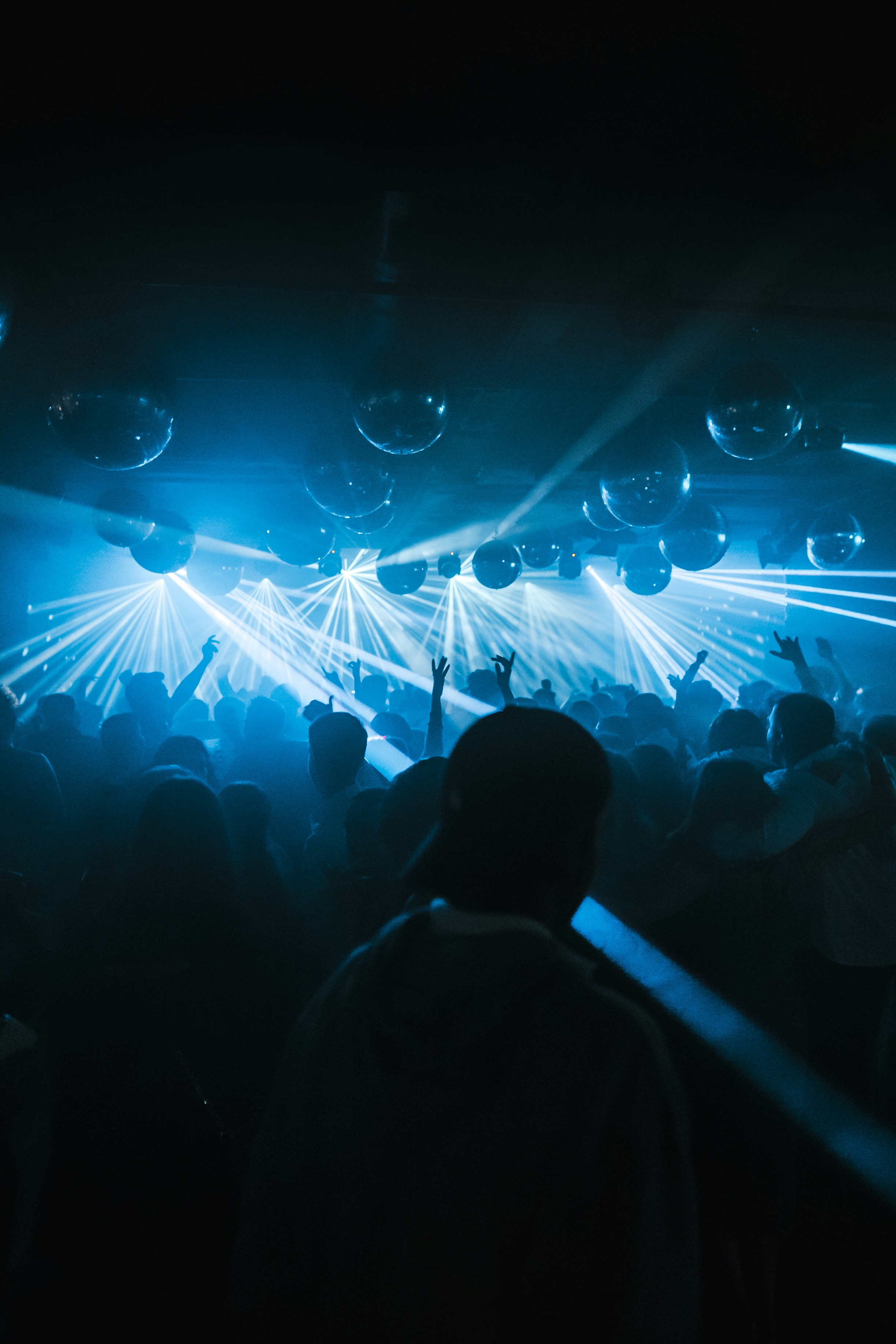}}

\begin{abstract}
Creativity support tools (CSTs) typically frame search as information retrieval, yet in practices like electronic dance music production, search serves as a creative medium for collage-style composition. To address this gap, we present \textsc{LoopLens}, a \added{research probe}\deleted{tool} for loop-based music composition that visualizes audio search results to support creative foraging and assembling. We evaluated \textsc{LoopLens} in a within-subject user study with 16 participants of diverse musical domain expertise, performing both open-ended (divergent) and goal-directed (convergent) tasks. Our results reveal a clear behavioral split: participants with domain expertise leveraged multimodal cues to quickly exploit a narrow set of loops, while those without domain knowledge relied primarily on audio impressions, engaging in broad exploration often constrained by limited musical vocabulary for query formulation. This behavioral dichotomy provides a new lens for understanding the balance between exploration and exploitation in creative search and offers clear design implications for supporting vocabulary-independent discovery in future CSTs.
\end{abstract}




\maketitle

\section{Introduction}

Creativity support tools (CSTs) have traditionally framed search as a means to an end, a way to retrieve information or assets that can then be consumed or incorporated into creative work~\cite{kules2005supporting}. 
However, in many creative domains, such as loop-based music creation, search serves a fundamentally different purpose: it is not merely a precursor to creativity, but also a creative medium in its own right. This distinction is particularly evident in electronic dance music (EDM) production, where artists engage in what we term ``search as creation'', using exploratory browsing and serendipitous discovery of audio loops as the primary compositional method. Unlike traditional music composition that begins with melodic or harmonic ideas, loop-based production treats search interfaces as creative canvases where artists assemble pre-existing elements into novel arrangements. The creative act emerges not from synthesis or generation, but from the curatorial process of discovery, selection, and juxtaposition --- a form of musical collage enabled by search.

\added{Professional digital audio workstations have recognized the importance of search by developing increasingly sophisticated capabilities. Modern tools such as Ableton Live 12~\cite{ableton} and Logic Pro~\cite{logicpro} offer tag-based semantic search, similarity-based recommendations, and tempo-matched previewing within their interfaces. Yet despite these advances, these tools continue to position search as auxiliary infrastructure ---  located in collapsible side panels and often temporally separated from the act of composition.  Moreover, while these search affordances have become more powerful, how they actually support the balance between exploration and exploitation that characterizes creative search, or how domain expertise shapes these search behaviors in creative contexts remains understudied. These kinds of questions are difficult to answer within professional DAWs, whose interface complexity and proprietary codebases make controlled study challenging.}
\deleted{Despite search being central ... that characterize creative search behaviors.}



\deleted{Recent advances in ... search behaviors in creative contexts.} To address these gaps, we present \textsc{LoopLens}, a \added{focused research probe}\deleted{tool} for loop-based\footnote{\added{Loops are short audio fragments that are ``generally associated with a single instrumental sound''~\cite{collins2013electronic}.}} music composition that \added{implements and extends common DAW search features in a controlled environment designed to investigate how search functions as a creative medium.}\deleted{treats search as a creative medium by visualizing audio search results through multiple complementary representations.loops are short audio fragments that are ``generally associated with a single instrumental sound''~\cite{collins2013electronic}.} \textsc{LoopLens} supports both broad exploratory discovery and targeted exploitative search through an integrated interface featuring switchable grid and list views, \deleted{novel} radial visualizations, and a ``Find Similar'' functionality that recommends both related and contrasting sounds, enabling users to engage with audio content through both auditory and visual modalities without requiring extensive musical vocabulary.

We investigate the following research questions: 
\begin{enumerate}[label=\textbf{RQ.\arabic*}]
    \item How do different creative goals (i.e., convergent vs. divergent tasks) affect users’ search strategies, composition processes, and interaction with creativity support tools? 
    \item How do users, with varying levels of musical expertise and divergent thinking abilities, utilize and perceive \added{underexplored}\deleted{novel} interface representations (e.g., radial waveforms) for different tasks?
    \item What are the roles and perceived utilities of optional exploratory and evaluative aids that provide feedback during the search and composition process? 
\end{enumerate}

We contribute a \added{mixed-methods }within-subject user study with 16 participants representing diverse levels of musical domain expertise, examining behavior across both open-ended (divergent) and goal-directed (convergent) creative tasks. Our empirical findings reveal a clear behavioral dichotomy based on expertise and task type: that convergent creative task elicits more querying, that participants with domain knowledge use more filters per query and treat composition as constraint satisfaction, and that participants with domain knowledge hesitate less before first insertion. We also observe uneven but positive uptake of radial waveforms and varied use of ``Find Similar'', a feature that recommends both the most similar and dissimilar sounds to support targeted search and serendipitous discovery. 


This research makes several contributions to the creativity support tools literature. First, we \added{examine}\deleted{introduce} the concept of ``search as creation'' and demonstrate its importance in loop-based music composition. Second, we provide empirical evidence for how domain expertise and task type shape creative search behavior, offering new insights into the exploration-exploitation trade-off in creative contexts. Finally, we derive concrete design implications for supporting vocabulary-independent discovery in future creativity support tools, particularly for domains where creative practice involves assemblage of existing materials rather than generating from thin air.

\section{Background \& Related Work}
\label{sec:related-work}

\subsection{Creativity \& Creativity Support Tool}

Creativity has been studied across many disciplines, leading to diverse definitions~\cite{shneiderman2006creativity}. We adopt Hewett et al.'s perspective~\cite{hewett2005creativity} that creativity can be considered a property of people, products, and a set of cognitive processes. This framing provides a well-scoped foundation for the design and evaluation of creativity support tools (CSTs), long considered a ``grand challenge'' in HCI~\cite{shneiderman2008creativity, remy2020evaluating}. 

\subsubsection{Models of Creative Thinking}
\label{sec:creative-thinking}

A central theme across models of creative thinking is the interplay between \textit{divergent} idea generation and \textit{convergent} evaluation~\cite{sowden2019shifting}. This dual-process perspective has inspired CST designs that scaffold flexible iteration in story writing~\cite{kim2025scaffolding} and empower users to balance exploration and exploitation in generative music systems~\cite{zhouInteractiveExplorationExploitationBalancing2021a}. However, much of HCI work on CSTs continues to emphasize idea generation~\cite{kim2025amuse,louie2020novice,louie2022expressive}, often using prompts that encourage producing new and potentially novel content (e.g., composing music to reflect the character and mood of an image~\cite{louie2020novice, huang2016chordripple} or writing a short chorus~\cite{kim2025amuse}). In contrast, our study introduces two tasks that differentially emphasize divergent and convergent thinking\added{ (\Cref{sec:method-tasks})}, enabling us to probe how CSTs might scaffold both modes of creativity and how search behaviors manifest within them.

\subsubsection{Evaluating CSTs}

Evaluation of CSTs remains heterogeneous, with no dominant framework~\cite{remy2020evaluating}. Following Hewett et al.~\cite{hewett2005creativity}, we adopt the perspective that creativity can be evaluated across people, products, and processes, and apply this framing to evaluate our own \deleted{system}\added{research probe} across these three dimensions\deleted{, with details in }\added{ (}\Cref{sec:method_evaluation}\added{)}. \Cref{tab:cst_evaluation} summarizes common approaches used in HCI research, such as self-report instruments (e.g., CSI~\cite{carroll2009creativity}), interaction logs~\cite{kim2025amuse},  expert judgment of creative outcomes~\cite{louie2022expressive}, homogenization analysis~\cite{anderson2024homogenization}, and creativity-related psychometrics (e.g., TTCT~\cite{torrance1968examples}, Gold-MSI~\cite{mullensiefen2014musicality}). For broader overviews of CST research, see Frich et al.~\cite{frich2019mapping} and Chung et al.~\cite{chung2021intersection}. 

\begin{table*}[!t]
    \centering
    \begin{tabular}{m{0.2\linewidth} m{0.35\linewidth} m{0.35\linewidth}}
        \toprule
       \textbf{Evaluating What?}  & \textbf{How?} & \textbf{Example(s)} \\ 
       \midrule
        Creative Process & Users' self-reported answers using questionnaires/survey & \shortstack[l]{NASA-TLX~\cite{hart1988development}, \\ Creativity Support Index (CSI)~\cite{carroll2009creativity}, \\  Cognitive Dimensions of Music Notations~\cite{nash2015cognitive}}\\
        Creative Process & Users' interaction-log & Kim et al.~\cite{kim2025amuse}  \\ 
        Creative Process & User interview & Most user studies do this \\ 
        \midrule
        Creative Outcome & Judgments by humans (experts and/or lay users)  &  Louie et al. ~\cite{louie2022expressive}, Loui et al.~\cite{loui2024sequence} \\ 
        Creative Outcome & Homogenization analysis & Anderson et al.~\cite{anderson2024homogenization}  \\ 
        \midrule
        \shortstack[l]{Creator Characteristics \\ (general-purpose)} & Torrance Test of Creative Thinking (TTCT)~\cite{torrance1968examples} & Loui et al.~\cite{loui2024sequence} \\
        \shortstack[l]{Creator Characteristics \\ (domain-specific)} & Music Divergent Thinking Test & van Welzen et al. ~\cite{van2024does} \\ 
        \shortstack[l]{Creator Characteristics \\ (prior training)}  & Goldsmith Musical Sophistication Index (Gold-MSI)~\cite{mullensiefen2014musicality} & Loui et al.~\cite{loui2024sequence} \\ 
        \shortstack[l]{Creator Characteristics \\ (reward sensitivity)}  & extended Barcelona Music Reward Questionnaire (eBMRQ)~\cite{cardona2022forgotten} & Loui et al.~\cite{loui2024sequence} \\ 
        \bottomrule \\
    \end{tabular}
    \caption{Common approaches to evaluating creativity support tools (CSTs) in HCI, organized by whether they target creative processes, outcomes, or creator characteristics. The table illustrates both widely used\added{ domain-general} instruments (e.g., CSI, TTCT) and domain-specific measures (e.g., Gold-MSI, Music Divergent Thinking Test).}
    \label{tab:cst_evaluation}
\end{table*}

\subsubsection{CST for Music Composition}


Recent music-related CSTs, such as COCOCO~\cite{louie2020novice}, AMUSE~\cite{kim2025amuse}, and MAICO~\cite{rauMAICOVisualizationDesign2025}, utilize symbolic music representations (e.g., MIDI, piano scroll view) and generative AI capabilities. Other studies have explored audio-based CSTs outside of music composition, such as Kamath et al.~\cite{kamathSoundDesignerGenerativeAI2024}, that looked at general sound effects / audio engineering tasks. Despite this progress, most music-related CSTs treat creativity primarily as generation from scratch. Yet, in many musical practices, particularly loop-based and electronic music, creativity also involves \textit{searching}, selecting, and combining from large sound libraries. While search has long been linked to creativity~\cite{kules2005supporting}, it is often externalized outside CST workflows. Our work foregrounds search as integral to loop-based music composition, extending perspectives such as Chavula et al.'s exploration of search for idea generation~\cite{chavula2023searchidea} to the domain of loop-based electronic music. 

\subsection{Electronic Dance Music\added{ and Musique Concr{\`e}te}}

We focus on Electronic Dance Music (EDM), a broad meta-genre including techno, house, trance, and drum `n' bass~\cite{butler2006unlocking}. EDM production is distinct in its reliance on raw audio rather than symbolic notation, using synthesizers, drum machines, sequencers, and samplers~\cite{butler2006unlocking, collins2013electronic}. Composition is often \textit{mosaic-like}: short loops and sequences are layered, repeated, and transformed, privileging cyclical repetition over linear progression. These features make EDM a particularly relevant context for studying how CSTs can integrate search and visual presentations of audio information.

\added{Beyond EDM, the practice of composing through assembling pre-recorded sounds has roots in \textit{musique concr{\`e}te}, which was introduced by Pierre Schaeffer in the 1940s~\cite{dack2013collage, battier2007grm, palombini1999musique}. While musique concr{\`e}te composers typically worked with environmental and instrumental sounds rather than repeatable musical loops, its compositional process similarly positions search and discovery as central creative acts rather than auxiliary steps.}

\subsection{Visualizing Musical Structure and Repetition}
\label{sec:vis-music-structure}

\begin{figure*}[!htbp]
    \centering
    \includegraphics[width=0.85\linewidth]{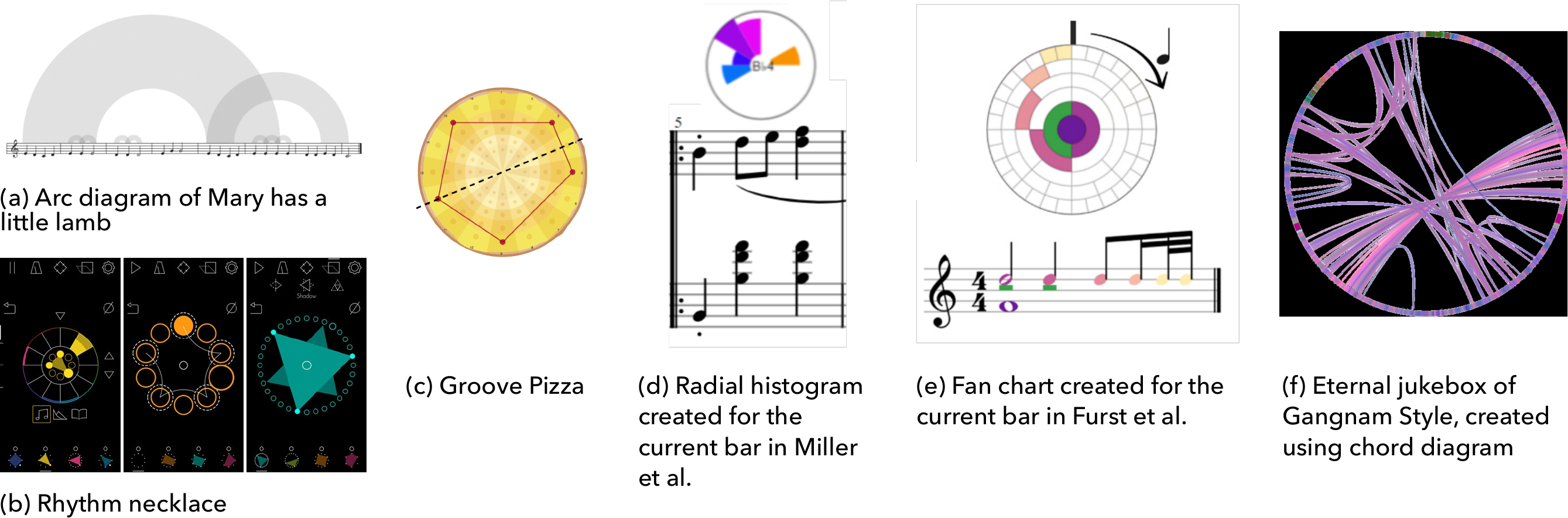}
    \caption{Example of existing radial charts and diagrams. Except the Eternal Jukebox (f), all other forms of visualizing music information takes in ``nice'' input formats that are not strictly audio but symbolic music notation.\added{ See Draper et al.~\cite{draper2009survey} for a survey on radial methods for visualization and Lima et al.~\cite{lima2021survey} for music visualization techniques.}}
    \label{fig:radial}
\end{figure*}

Standard audio visualizations such as waveforms and spectrograms plot amplitude or frequency \textit{linearly} over time. To better represent music's cyclic and repetitive nature, researchers\deleted{ and scholars} have explored \textit{radial} layouts, from medieval music theory diagrams~\cite{wright1995preliminary} to modern arc diagrams~\cite{wattenberg2002arc}, radial ``fingerprints'' for harmony\added{ and rhythm}~\cite{miller2019augmenting, furst2020augmenting}, and spherical projections~\cite{lopez2019music}. Interactive tools such as the \textit{Eternal Jukebox}~\cite{eternalJukebox}, \textit{Groove Pizza}~\cite{groovePizza}, and \textit{Rhythm Necklace} help users explore loops and rhythms via circular layouts \added{(\Cref{fig:radial})}. \added{In the professional domain, Apple's Logic Pro utilizes radial waveforms within its \texttt{Live Loops} grid \includegraphics[height=1em]{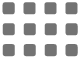}, which is ``a grid of musical loops and phrases that [users] can trigger and manipulate in real time to create unique arrangements''~\cite{logicpro}. Here, the compact radial shape serves a \textit{spatial} purpose: it allows waveforms to fit neatly into a dense grid of cells, mirroring the layout of hardware controllers (e.g., Launchpad) used for live performance.} 

\added{Our work also uses radial waveform visualization, but instead of using it during the arrangement phase, we hypothesize that the compact nature of radial waveforms makes them easy for users to scan and compare during the search phase, a task that remains difficult with traditional linear waveforms.}\deleted{However, most approaches rely on ... for music visualization techniques.}


\subsection{\added{Search Interfaces in Commercial Digital Audio Workstations}}
\label{sec:search-interface}

\begin{figure}
    \centering
    \includegraphics[width=0.95\linewidth]{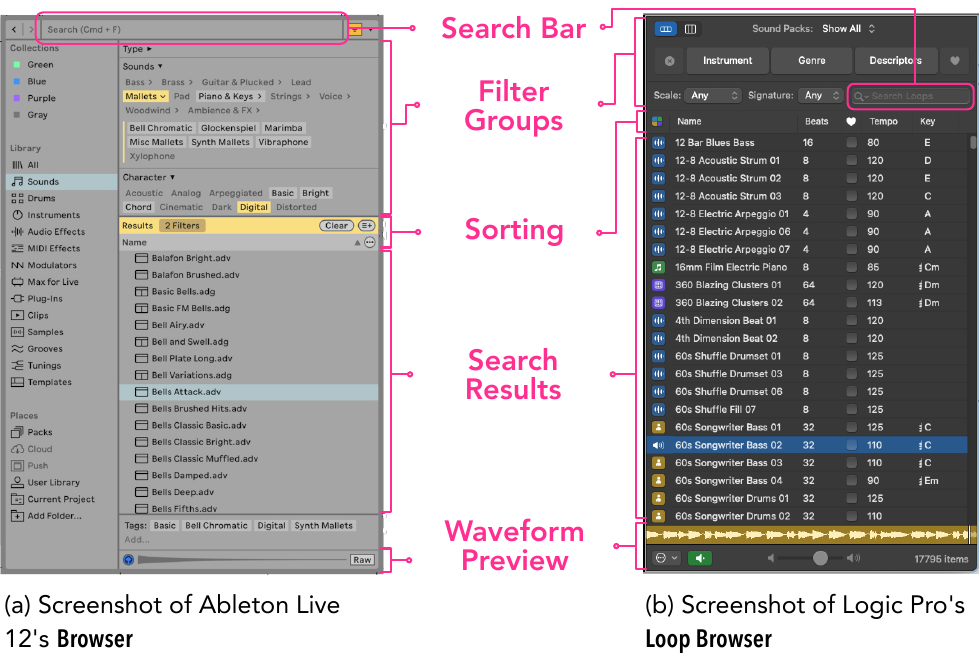}
    \caption{\added{Screenshots from Ableton Live 12 and Apple Logic Pro's interfaces for searching and browsing music.}}
    \label{fig:daw-overall}
\end{figure}

\added{To situate our work within current music production practice, we examined two industry-standard Digital Audio Workstations (DAWs)~\cite{bestdaw}: \textbf{Ableton Live 12} (released March 2024~\cite{abletonTwelveRelease}) and \textbf{Logic Pro} 11.2 (released May 2025). Both tools offer sophisticated search capabilities. In Ableton Live 12's \textbf{Browser}, in addition to standard keyword search and hierarchical folder navigation, users can filter by tag-based semantics (e.g., ``bright'', ``aggressive''), access a ``Similar Sound'' feature that surfaces related sounds, and preview loops with automatic tempo-matching to the main project. Logic Pro offers comparable functionality through its \textbf{Loop Browser} (\cref{fig:daw-overall} (b)), including tag-based filtering and tempo synchronization. For composition, both support drag-and-drop interactions from the browser to the timeline, enabling easy transitions from search to arrangement.}

\added{We observe that the interface design of these browsers reflects a priority on composition over discovery. The search panels are collapsible and list-based, optimizing screen space for arrangement while implicitly framing search as a transient, auxiliary task. While this file-system rooted list interface supports professionals' need for managing vast sound samples~\cite{andersen2016conversations}, it imposes a high cognitive load on novices. For example, distinguishing between generic titles like \textit{60s Songwriter Bass 02} and \textit{60s Songwriter Bass 03} that share the same key and tempo requires users to audition samples serially to discern sonic differences, as the interface provides no visual cues (\cref{fig:daw-overall} (b)). In Logic Pro, waveform previews only appear during playback, further preventing visual comparison of similar loops. Investigating these affordances within a commercial DAW is also challenging: their proprietary nature prevents easy modifications of existing interface parameters, and their steep learning curves act as confounding variables.}

\section{Design Principles}

Based on gaps identified in prior literature\added{ and features observed in existing industry-standard DAWs (\Cref{sec:search-interface})}, we \added{developed}\deleted{designed} \textsc{LoopLens}\added{ as a \textbf{research probe}. Our goal was to create a controlled environment that retains core DAW functionalities while isolating specific search behaviors from the confounding complexity of full-scale production software. The following design principles guided our approach}\deleted{with the following design principles in mind}: 
\begin{enumerate}[label={\bfseries DP.\arabic* }, leftmargin=*]
    \item \textbf{\added{Prioritize discovery over retrieval}\deleted{Treat search as integral to creation}}.\added{ Rather than treating search as a collapsible auxiliary to composition, we foreground the search browser while making the sequencer grid toggle-able}\deleted{Traditional DAWs separate the process of finding sounds (browsing folders) from using them (arranging on a timeline). We aimed to dissolve this boundary}. This principle is realized in \textsc{LoopLens} through the integrated browser and DAW, where users can right-click a loop within their composition to immediately find and audition similar sounds in context.
    \item \textbf{Support \added{multi-scale visual comparison}\deleted{multiple representational view}}. Different tasks require different views of the same information. To support both high-level \deleted{visual} browsing and detailed comparison, our \added{search results section}\deleted{browser} features both a grid view, inspired by Tufte's ``small multiples''~\cite{tufte1991envisioning}, and a traditional list view\added{ commonly seen in existing DAWs (\cref{fig:daw-overall})}. This allows users to switch between modes depending on their current goal, from exploration to focused selection.
    \item \textbf{Scaffold creative workflows without over-automation}.\added{ Research on constrained digital instruments suggests that constraints are often more effective than open-ended possibilities in facilitating the emergence of personal style~\cite{gurevich2012playing}. } While the goal is to support both experts and novices, the tool should not overly constrain them or make creative decisions for them. \textsc{LoopLens} provides structure, such as the beat-gridded timeline and the ``Find similar'' feature, but leaves all compositional choices, such as layering, arrangement, and final selection, entirely up to the user. This provides a ``low threshold'' for getting started without imposing a ``low ceiling'' on creative possibilities~\cite{myers2000past}.
\end{enumerate}

\begin{figure*}[htbp]
    \centering
    \includegraphics[width=1\linewidth]{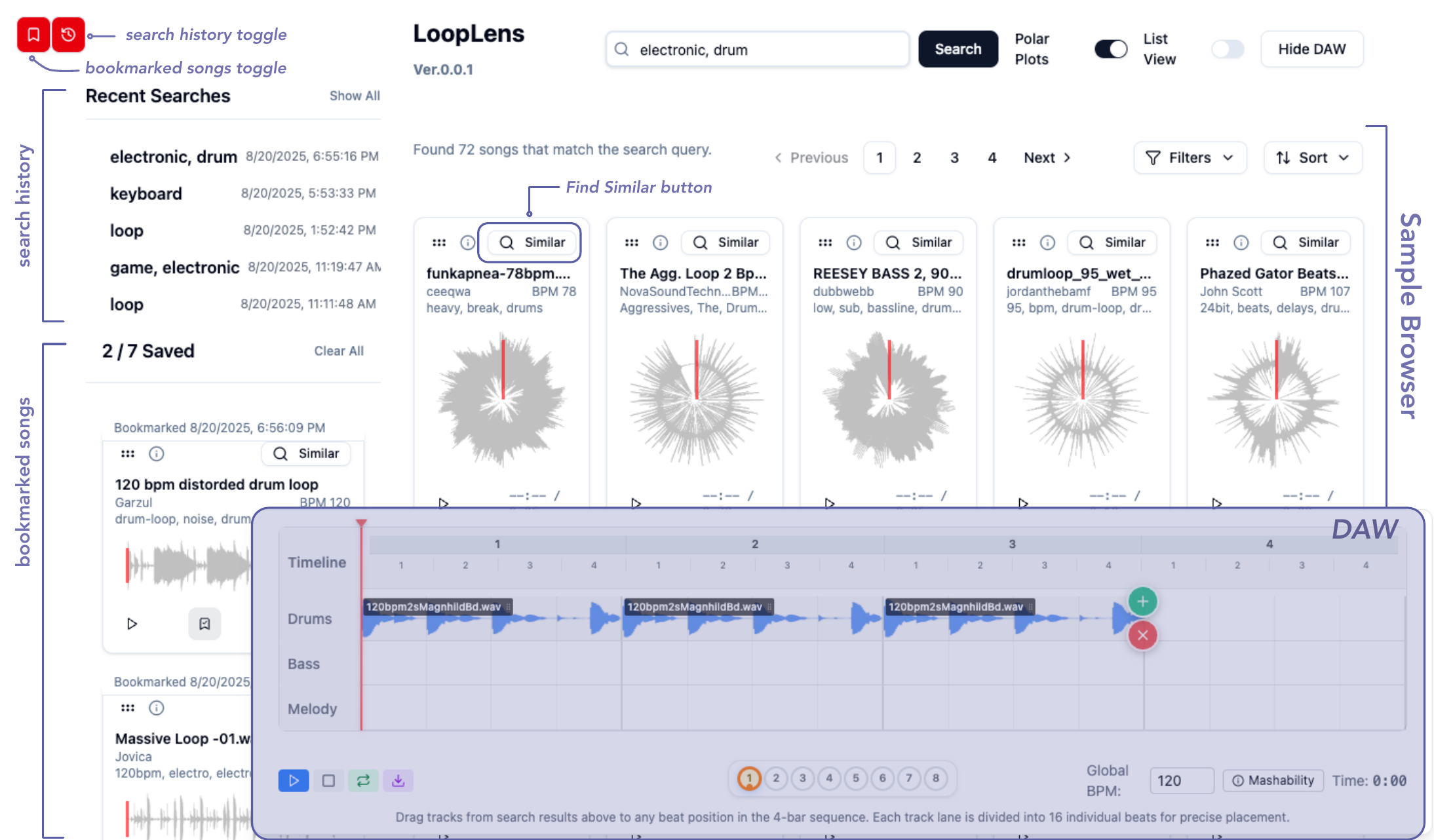}
    \caption{Screenshot of \textsc{LoopLens} user interface.\added{ Note that the purple shade for DAW is manual annotation and not what the actual interface appears like.}}
    \Description{Interface}
    \label{fig:interface}
\end{figure*}

\section{\textsc{LoopLens}: System Overview}

\textsc{LoopLens} is \added{a research probe implemented as }an interactive web application designed for exploring and composing with a database of musical loops. The system integrates three core elements: a \textbf{sample browser} for searching and comparing loops, an embedded \textbf{digital audio workstation (DAW)} for arranging compositions, and lightweight \textbf{discovery tools} such as the ``Find Similar'' modal that surface related or contrasting sounds. These components are tightly coupled to support a fluid workflow: users can search and/or discover candidate loops, preview them in different visual forms, and drag and drop them directly into the DAW timeline for composition. 

\subsection{Interface Overview}

\begin{figure}[htbp]
  \begin{center}
    \includegraphics[width=0.5\linewidth]{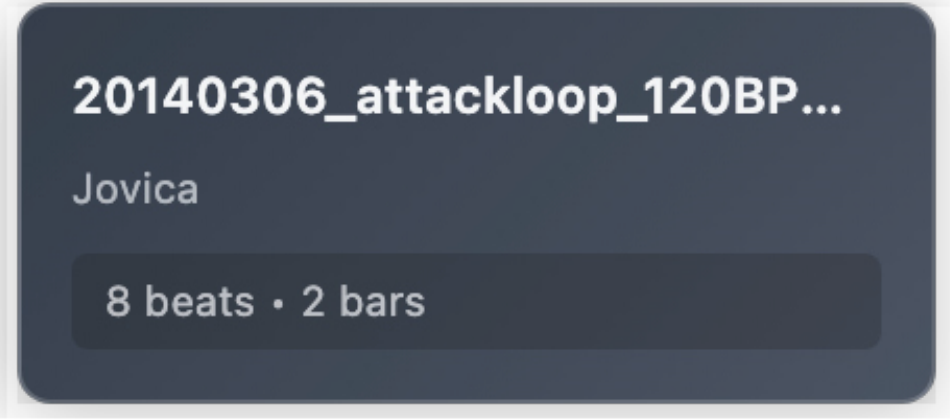}
  \end{center}
  \caption{\added{Preview of song during drag and drop.}}
  \label{fig:drag-preview}
\end{figure}

The sample browser (\Cref{sec:sample-browser}) and the DAW (\Cref{sec:daw}) are tightly integrated to support a seamless workflow. Once a user finds a desired loop, they can \textit{drag and drop} it directly from the browser into the DAW timeline. Furthermore, users can bookmark loops for later consideration and subsequently drag them from their bookmarked collection into the composition. For serendipitous discovery, users can also explore similar loops through the ``Find Similar'' modal (\Cref{sec:find_similar}) and drag these directly into the timeline. During the drag-and-drop process, the interface displays how many beats the loop contains and how many bars it will occupy in the timeline\added{ (\cref{fig:drag-preview})}. 

\subsubsection{Sample Browser}
\label{sec:sample-browser}

To accommodate different workflows, users can toggle between two visualization modes (``linear plots'' vs ``polar plots'') for the loop database. The \textit{grid view} presents loops as ``small multiples''~\cite{tufte1991envisioning}, facilitating rapid visual comparison and enabling spatial browsing across rows and columns, rather than a strictly top-to-bottom sequence. This contrasts with the \textit{list view}, a more traditional layout common in commercial sample libraries like \textit{Splice}~\cite{splice} and \textit{LANDR}~\cite{landr}, which presents data in a dense, tabular format. To facilitate discovery, the browser includes a keyword search that queries loop titles and tags. \added{Similar to existing commercial DAWs, }users can further refine the results using filters for BPM, key, and instrument, and sort the filtered view by relevance or other metadata attributes. This combination of search, filtering, and sorting allows users to move from broad exploration to specific selection efficiently.

\subsubsection{Digital Audio Workstation (DAW)}
\label{sec:daw}

\begin{figure}[htbp]
    \centering
    \includegraphics[width=\linewidth]{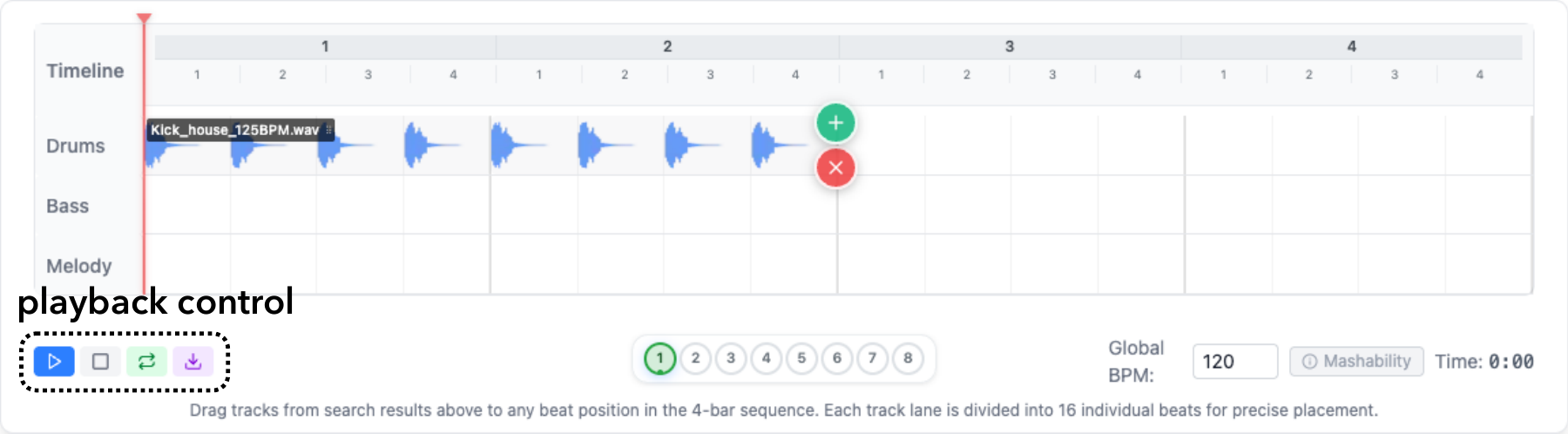}
    \caption{Screenshot of DAW in \textsc{LoopLens}.}
    \label{fig:daw}
\end{figure}

The integrated DAW replicates the core functionalities of standard music production software. At the global level, users can manage playback and adjust the project's beats per minute (BPM).\added{ Users can also adjust the volume for each of the three tracks (Drums, Bass, and Melody).} The timeline is based on the Time Unit Box System~\cite{toussaint2004geometry}, where each grid cell represents one beat. The duration of a beat is determined by the global BPM (e.g., at 120 BPM, each beat is 0.5 seconds). Each page displays four bars in a 4/4 time signature, a structure that aligns with the most common rhythmic pattern (i.e., ``four on the floor'') found in electronic music. At the track level, users can duplicate or remove loops and right-click to access functions like bookmarking or finding similar sounds directly within the composition context.

\subsubsection{Find Similar Modal}
\label{sec:find_similar}

\begin{figure}[htbp]
  \begin{center}
    \includegraphics[width=\linewidth]{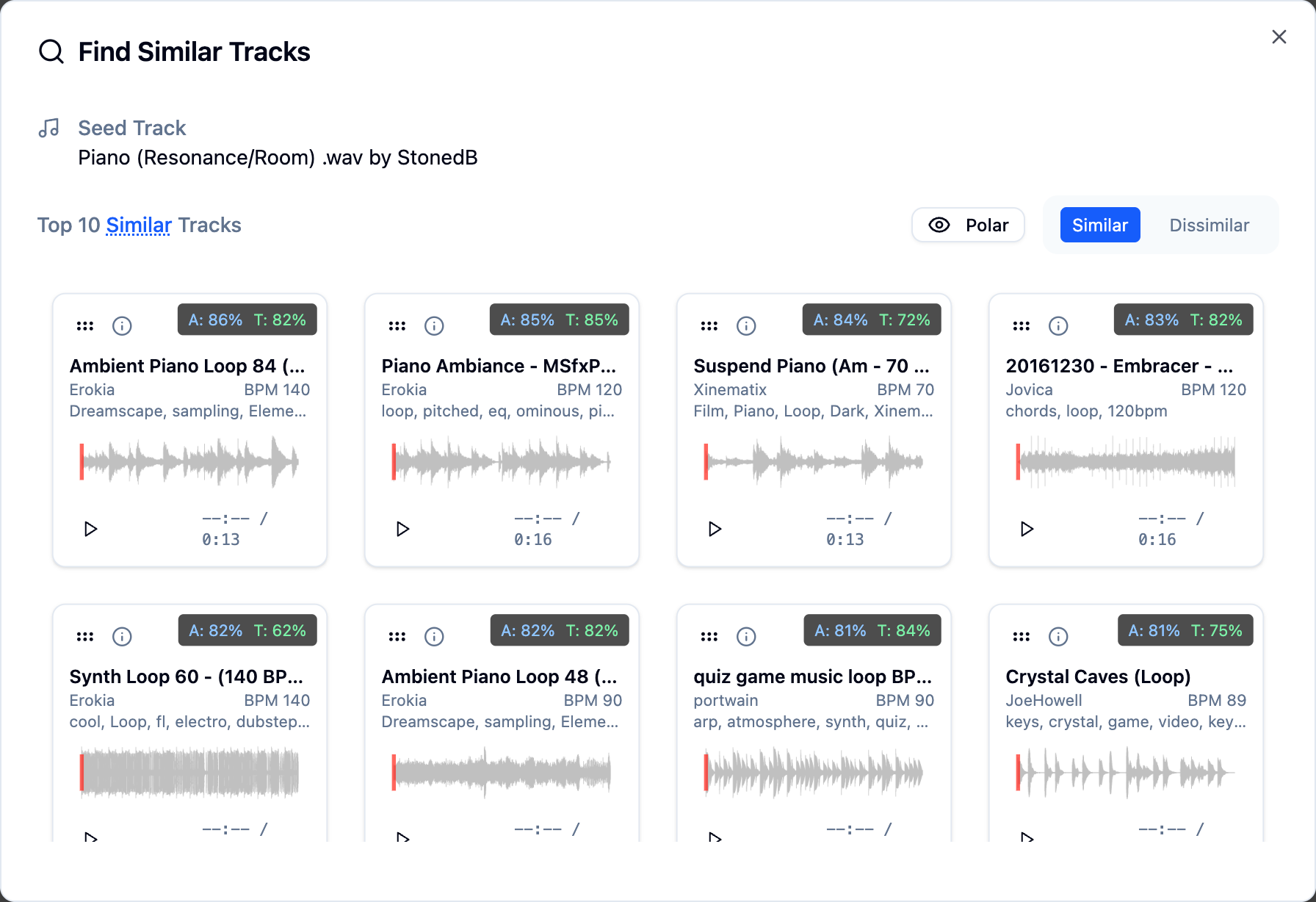}
  \end{center}
  \caption{Screenshot of ``Find Similar'' modal for the song, \textit{Piano (Resonance/Room) .wav}, by StonedB, in \textsc{LoopLens}.}
\end{figure}

To support discovery beyond keyword search and filtering, \textsc{LoopLens} provides a ``Find Similar'' feature. For each loop, users can click the ``Similar'' button to open a modal that recommends related sounds. The system retrieves the 10 most similar loops based on deep audio embeddings computed with CLAP~\cite{wu2023large}, as well as the 10 most dissimilar loops to encourage contrastive exploration.\added{ We include the dissimilar sounds because Andersen and Knees~\cite{andersen2016conversations} mentions it an interesting challenge for creative Music Information Retrieval Systems to solve.} Results in the modal adopt the same dual visualization options as the main browser (i.e., linear or polar waveforms), ensuring consistency across the interface. 

Like other parts of the system, loops shown in the modal are fully interactive: users can preview them, bookmark them, or drag them directly into the DAW timeline. This design positions the modal as a lightweight recommendation tool that supports both \textit{targeted search} (finding sounds that blend well with a current loop) and \textit{serendipitous discovery} (surfacing contrasting material that might inspire variation).



\subsection{Implementation}

\textsc{LoopLens} is a web application built using \texttt{SvelteKit}, a full-stack framework. We implemented audio-related functionalities using \texttt{Tone.js} (\url{https://tonejs.github.io/}), a Web Audio API framework for creating interactive music in the browser. For the loop database, we utilized the Freesound Loop Dataset~\cite{ramires2020freesound}, a large-scale dataset of annotated music loops. The loops were collected from Freesound~\cite{font2013freesound} (\url{https://freesound.org/}), a community database of audio recordings released under Creative Commons license, allowing for reproduction. The original dataset contains 9040 loop sounds with their metadata, and we further filtered this dataset to ensure the validity of the annotated labels of beats per minute (bpm), key, and instruments, resulting in a final annotated dataset of \textbf{331} loops, which form the basis of our loop database to be used for composition. Given the limited size of the database, we implemented filters disjunctively (OR), so a loop satisfying any active constraint can appear in the results. The radial and linear waveform visualizations were generated using the \texttt{librosa}~\cite{mcfee2015librosa} and \texttt{matplotlib}~\cite{Hunter:2007} libraries. 

\subsubsection{Find Similar Song}

Musical similarity is multi-dimensional, context-based, and hard to define~\cite{rocha2013segmentation, panteli2017model}. As it is beyond the central aim of this paper to develop better similarity measures of music, we opted to use deep features obtained from Contrastive Language-Audio Pretraining (CLAP)~\cite{wu2023large} for calculating distances between audio embeddings and text embeddings of concatenated tags as a way to operationalize the calculation of musical similarity. Specifically, we employ the \href{https://huggingface.co/laion/clap-htsat-unfused}{\texttt{laion/clap-htsat-unfused}} model, which uses a Hierarchical Token Semantic Audio Transformer (HTSAT) architecture to generate fixed-length audio embeddings from raw waveforms. The unfused variant enables separate generation of audio and text embeddings, allowing independent similarity scoring for each modality. Pairwise similarity between tracks is computed using cosine similarity between audio embeddings, which ranged from $-1$ to $1$, which are then normalized to a $0-1$ scale for user interface display. We provide the top 10 most similar and most dissimilar tracks by audio similarity. 

\section{Methodology}

\subsection{Experiment Procedure}

\begin{figure}[htbp]
  \begin{center}
    \includegraphics[width=0.48\textwidth]{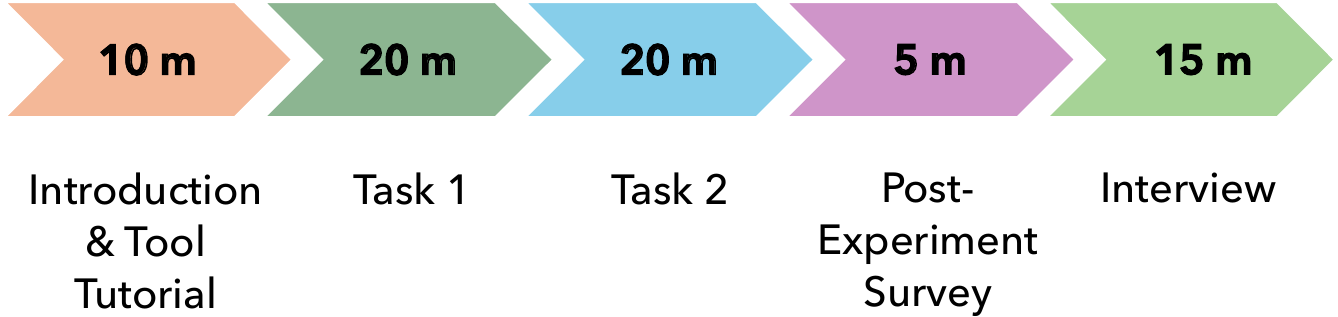}
  \end{center}
  \caption{Overview of the user study procedure.}
  \label{fig:procedure}
\end{figure}

\Cref{fig:procedure} provides an overview of the overall study procedure. Prior to the study session, participants completed a pre-experiment survey to capture their prior musical training, such as level of experience and involvement with music and exposure to DAW. Similar to Loui et al.~\cite{loui2024sequence}, we selected questions from the \textit{Goldsmiths Musical Sophistication Index} (Goldsmiths MSI) questionnaire~\cite{mullensiefen2014musicality} that are relevant to musical training. Participants were then directed to complete a general-purpose divergent thinking task, i.e., the \textit{Divergent Association Task} (DAT) developed by Olson et al.~\cite{olson2021naming}. In this task, participants are asked to generate a set of ten unrelated words, and their semantic distance is scored computationally. We selected the DAT because it provides a domain-general measure of cognitive flexibility that can be administered quickly, making it suitable for studies where creative capacity is of interest but time is limited. 

During the study, participants were first given 10 minutes of guided instructions to familiarize themselves with the interface, and then went through a session composed of two tasks, each taking 25 minutes. The order of the tasks were counterbalanced to minimize effects of the ordering bias. After participants finished the tasks, they were directed to complete a post-experiment survey, consisted of eight 7-point Likert-scale questions, before a 15 minute semi-structured interview. The whole process was audio and screen-recorded via Microsoft Teams. Details of the survey and interview questions can be found in \Cref{sec:survey,sec:interview}. 

\subsection{Participants}

We recruited 16 participants (5 women, 11 men, ages 23 -- 49, mean = 30.37, SD = 8.69) from employees at a large multinational corporation through internal communication channels. Due to anonymization policy and confidentiality agreements, we cannot disclose the specific organization name. All participants currently reside in Tokyo, Japan but represent a range of linguistic and cultural backgrounds. The study was conducted in English, which was not the first language for most participants, though it is one of the primary working languages of the organization. Occasional Japanese was used for clarification. 

We acknowledge that single-organization recruitment may introduce organizational bias, which we addressed by targeting employees across different departments and roles, making the participation entirely voluntary, and ensuring that no identifying information was shared with the employer. Participants also varied in musical experience and prior exposure to DAW, further broadening the diversity of perspectives in the study. Details related to participant's background can be found in~\Cref{tab:participant}. 

\begin{table*}[!htbp]
    \centering
    \begin{tabular}{m{2cm} m{4cm} m{2.5cm} m{3.5cm}}
\toprule

{\bfseries Participant ID} & {\bfseries Prior Musical Training} & {\bfseries DAW} & {\bfseries Company Role} \\ 

\midrule\addlinespace[2.5pt]

\texttt{P1} & 3/2/0 & {\cellcolor[HTML]{3182BD}{\textcolor[HTML]{FFFFFF}{Advanced}}} & {\cellcolor[HTML]{DECBE4}{\textcolor[HTML]{000000}{Research}}} \\ 

\texttt{P2} & 0/6-9/6-9 & {\cellcolor[HTML]{EFF3FF}{\textcolor[HTML]{000000}{None}}} & {\cellcolor[HTML]{DECBE4}{\textcolor[HTML]{000000}{Research}}} \\ 

\texttt{P3} & 1/3-5/0 & {\cellcolor[HTML]{BDD7E7}{\textcolor[HTML]{000000}{Beginner}}} & {\cellcolor[HTML]{FBB4AE}{\textcolor[HTML]{000000}{Engineering}}} \\ 

\texttt{P4} & 4-6/6-9/10+ & {\cellcolor[HTML]{3182BD}{\textcolor[HTML]{FFFFFF}{Advanced}}} & {\cellcolor[HTML]{DECBE4}{\textcolor[HTML]{000000}{Research}}} \\ 

\texttt{P5} & 0.5/0/0 & {\cellcolor[HTML]{3182BD}{\textcolor[HTML]{FFFFFF}{Advanced}}} & {\cellcolor[HTML]{DECBE4}{\textcolor[HTML]{000000}{Research}}} \\ 

\texttt{P6} & 0/3-5/10+ & {\cellcolor[HTML]{EFF3FF}{\textcolor[HTML]{000000}{None}}} & {\cellcolor[HTML]{DECBE4}{\textcolor[HTML]{000000}{Research}}} \\ 

\texttt{P7} & 4-6/10+/1 & {\cellcolor[HTML]{6BAED6}{\textcolor[HTML]{000000}{Intermediate}}} & {\cellcolor[HTML]{DECBE4}{\textcolor[HTML]{000000}{Research}}} \\ 

\texttt{P8} & 4-6/0/0 & {\cellcolor[HTML]{3182BD}{\textcolor[HTML]{FFFFFF}{Advanced}}} & {\cellcolor[HTML]{DECBE4}{\textcolor[HTML]{000000}{Research}}} \\ 

\texttt{P9} & 0/0/0 & {\cellcolor[HTML]{EFF3FF}{\textcolor[HTML]{000000}{None}}} & {\cellcolor[HTML]{FBB4AE}{\textcolor[HTML]{000000}{Engineering}}} \\ 

\texttt{P10} & 3/6-9/6-9 & {\cellcolor[HTML]{EFF3FF}{\textcolor[HTML]{000000}{None}}} & {\cellcolor[HTML]{FBB4AE}{\textcolor[HTML]{000000}{Engineering}}} \\ 

\texttt{P11} & 0/0/10+ & {\cellcolor[HTML]{EFF3FF}{\textcolor[HTML]{000000}{None}}} & {\cellcolor[HTML]{DECBE4}{\textcolor[HTML]{000000}{Research}}} \\ 

\texttt{P12} & 1/3-5/0 & {\cellcolor[HTML]{6BAED6}{\textcolor[HTML]{000000}{Intermediate}}} & {\cellcolor[HTML]{DECBE4}{\textcolor[HTML]{000000}{Research}}} \\ 

\texttt{P13} & 0/3-5/3 & {\cellcolor[HTML]{EFF3FF}{\textcolor[HTML]{000000}{None}}} & {\cellcolor[HTML]{FBB4AE}{\textcolor[HTML]{000000}{Engineering}}} \\ 

\texttt{P14} & 0/0/0 & {\cellcolor[HTML]{EFF3FF}{\textcolor[HTML]{000000}{None}}} & {\cellcolor[HTML]{FED9A6}{\textcolor[HTML]{000000}{Product Management}}} \\ 

\texttt{P15} & 1/10+/10+ & {\cellcolor[HTML]{EFF3FF}{\textcolor[HTML]{000000}{None}}} & {\cellcolor[HTML]{FBB4AE}{\textcolor[HTML]{000000}{Engineering}}} \\ 

\texttt{P16} & 0/0.5/10+ & {\cellcolor[HTML]{EFF3FF}{\textcolor[HTML]{000000}{None}}} & {\cellcolor[HTML]{FBB4AE}{\textcolor[HTML]{000000}{Engineering}}} \\ 

\bottomrule \\

\end{tabular}
    \caption{Overview of details related to participant's backgrounds. For ``prior musical training'', the numbers correspond to the number of years received in musical theory training, music instrument training, and total years of practicing music \added{on a daily and regular basis, corresponding to questions (3), (4), and (1) in \Cref{sec:survey}. A participant could have no training in music theory, yet could have been practicing music on a daily basis for more than 10 years (e.g., \texttt{P16})}\deleted{respectively}.}
    \label{tab:participant}
\end{table*}

\subsection{Tasks}
\label{sec:method-tasks}

\begin{figure}[htbp]
  \begin{center}
    \includegraphics[width=0.6\linewidth]{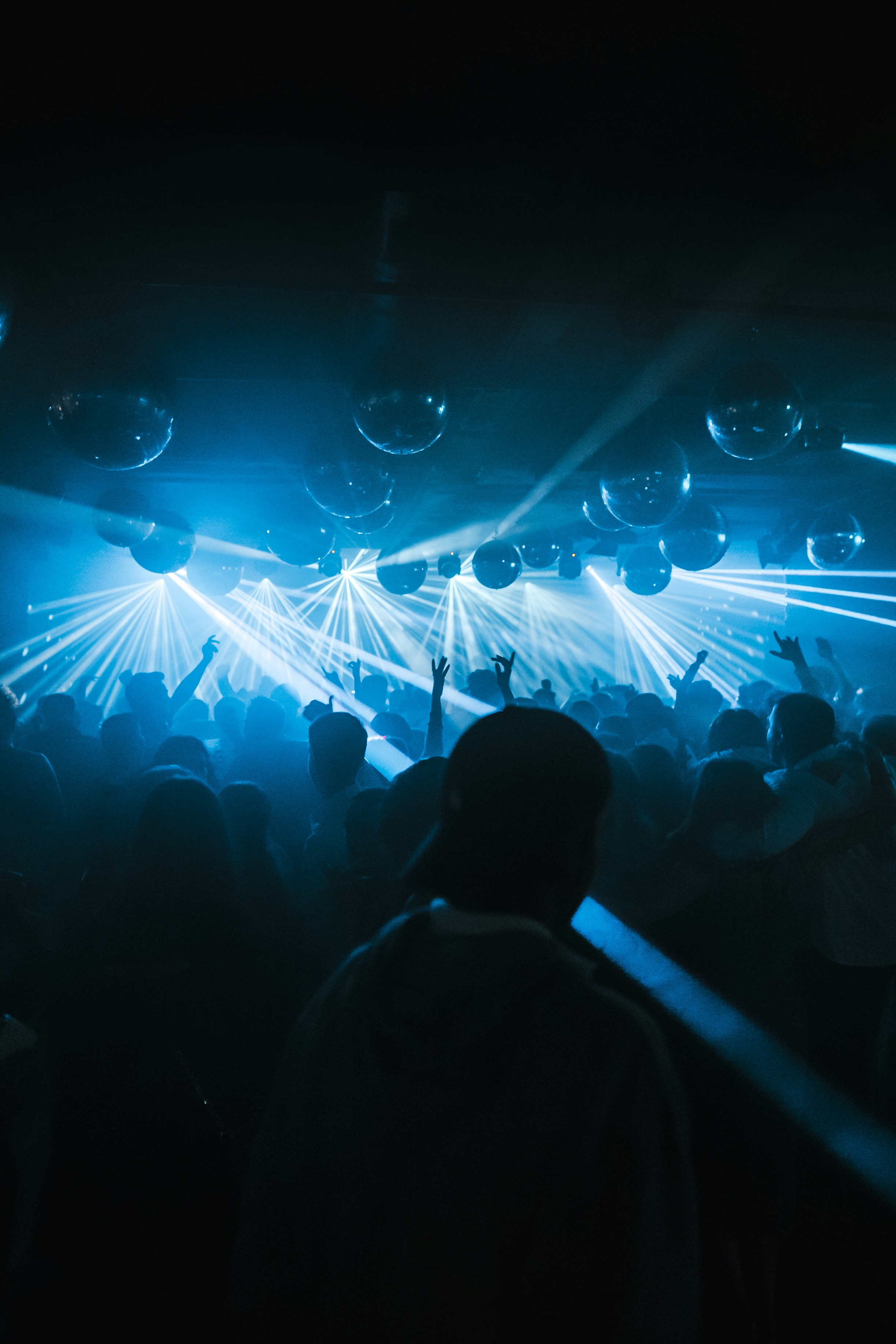}
  \end{center}
  \caption{Picture prompt for divergent task.}
  \label{fig:div-image}
  \Description{A picture with blue background light showing people watching concert during night time.}
\end{figure}

As discussed in \Cref{sec:creative-thinking}, we employed two different tasks, one more convergent, the other more divergent. For both tasks, we asked the user to compose a music piece of at least 16 bars in length. 
These task designs were chosen to systematically examine different modes of creative thinking in musical composition. 

For the \includegraphics[height=1em]{img/convergent.pdf} convergent task, we asked the user to listen to a sample music of 32 seconds and create a music piece that captures \textbf{\textit{the same rhythmic feel and energy}} as the sample. Users were allowed to freely use all the information, including audio, \added{visual} waveform, as well as tags provided by the original composer, and were allowed to revisit these information as often as needed. The sample music, titled ``\texttt{Arcade Music Loop.wav}'', was obtained from Freesound at \url{https://freesound.org/people/joshuaempyre/sounds/251461/}. The convergent task represents a constraint-based creative process where participants work toward a specific, identifiable target, and thus it provides a more bounded creative space with measurable success criteria, similar to how convergent thinking seeks optimal solutions within defined parameters. 

For the \includegraphics[height=1em]{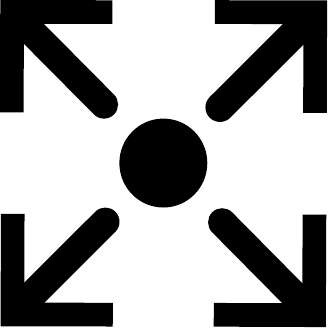} divergent task, similar to prior work~\cite{kim2025amuse,louie2020novice,louie2022expressive}, we asked the user to create a musical piece that \textbf{\textit{matches the mood}} of the shown image in~\Cref{fig:div-image}. The scene of this visual stimuli could be interpreted across multiple creative dimensions, not only musical elements like tempo and harmony, but also environmental audio characteristics such as ambient crowd noise. This multiplicity of valid interpretational pathways, with no single ``correct'' solution, aligns with divergent thinking's emphasis on generating multiple novel possibilities from a single prompt.



\subsection{Evaluation}
\label{sec:method_evaluation}

We use \texttt{tidyverse}~\cite{tidyverse}, \texttt{reticulate}~\cite{reticulate}, and \texttt{jsonlite}~\cite{jsonlite} for data analysis.   We use \texttt{ggplot2}~\cite{ggplot2}, \texttt{ggbeeswarm}~\cite{ggbeeswarm}, \texttt{wordcloud2}~\cite{wordcloud2} for plotting results in the paper. Pre-trained weights were obtained from \texttt{HuggingFace}. 

\subsubsection{Creator Characteristics}

\begin{figure*}[htbp]
        \centering
    \includegraphics[width=0.75\linewidth]{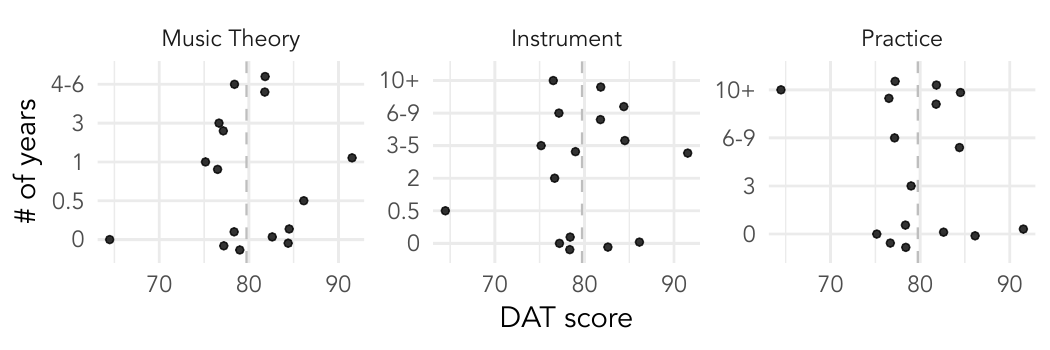}
    \caption{Scatter plots showing the relationship between participant DAT scores and self-reported years of experience, faceted by three training types: Music Theory, Instrument, and Practice. The vertical dashed line represents the mean DAT score.}
    \label{fig:dat-score}
\end{figure*}

To quantify divergent thinking, we calculated a Divergent Association Task (DAT) score for each participant based on the 10 words they generated in the pre-study survey. Following the original methodology~\cite{olson2021naming} and using the authors' publicly available code\footnote{\url{https://github.com/jayolson/divergent-association-task}}, we computed the average semantic distance between all pairs of words in each participant's list. In general, higher DAT scores indicate greater conceptual divergence, as demonstrated by Olson et al.~\cite{olson2021naming}.

Our 16 participants achieved an average DAT score of $79.73$ (SD $5.96$), which closely aligns with the average of $78.38$ ($SD = 6.35$) reported by Olson et al~\cite{olson2021naming}. As shown in \Cref{fig:dat-score}, we observe  no clear linear relationship between years of experience and participant's ability to divergently think, as measured by DAT score. This suggests that while musical training builds domain-specific expertise, its effects may not directly translate to performance on a domain-general measure of creativity like the DAT. 

For comparative analysis in~\Cref{sec:results_process}, we categorized participants into two groups: \faBaby{}\textbf{Novices}, defined as those with no prior music theory training \textit{and} no DAW experience (7 out of 16 participants), and \faGlasses{} \textbf{Experts}, comprising all remaining participants.

\subsubsection{Creative Process}

We qualitatively analyzed data from two sources: semi-structured interviews and screen recordings. Two authors independently analyzed the interview transcripts using reflexive thematic analysis~\cite{braun2006using} to identify key themes in participants' self-reported experiences. In parallel, we performed behavioral coding of the screen recordings using Datavyu software~\cite{Datavyu2014} to capture participants' observed interactions with the composition interface. We developed a structured coding scheme that manually tagged key compositional behaviors (e.g., playback events, editing actions, reference material consultation), producing detailed timelines of each participant's creative process. Findings from both qualitative datasets were triangulated to provide a comprehensive understanding of how participants approached each task. 

For quantitative analysis of process measures, we used non-parametric tests due to our sample size. We analyzed participants' self-reported responses related to the within-subject factor (task type) using the Wilcoxon signed-rank test, and examined between-subject differences (novice vs. expert) using the Mann-Whitney U test.


\subsubsection{Creative outcome} 

Following Anderson et al.~\cite{anderson2024homogenization}, who adapted automated verbal creativity scoring methods\cite{beaty2021automating} to perform a homogenization analysis, we applied a similar approach to \textit{automatically score musical creativity} by calculating ``semantic'' distances in embedding space. For the divergent task, we computed the cosine similarity between each participant's composition and the average embedding of all compositions in that condition. This distance from the group average serves as a proxy measure for creative divergence, i.e., compositions further from the center representing more unique creative solutions. For the convergent task, we temporally aligned participant compositions with the 32-second reference piece using time-stretching to normalize for BPM differences, then calculated two similarity measures: (1) participant composition to reference music (measuring constraint adherence), and (2) participant composition to the group mean (measuring individual distinctiveness within the constraint). This dual approach allows us to evaluate both how well participants met the convergent task requirements and how much individual variation existed within those constraints.

To ensure robustness across different audio representation methods, we conducted a multiverse analysis~\cite{steegen2016increasing} using four pre-trained models: \texttt{CLAP}~\cite{wu2023large} \texttt{htsat-unfused}, \texttt{CLAP}~\cite{wu2023large} \texttt{larger\_clap\_music}, \texttt{MAP-MERT-v1}~\cite{li2023mert}, and \texttt{MAP-Music2Vec}~\cite{mccallum2022supervised}. These models employ contrastive learning and self-supervised training on large-scale audio datasets, representing different but theoretically justified approaches to extracting meaningful audio features and quantifying musical similarity. We tested our primary hypothesis --- that creative divergence (distance from group average) varies as a function of participants' musical expertise, divergent thinking abilities, and task type --- across all four embedding spaces. We report results from all models to provide transparent and comprehensive findings.

\section{Results}
\label{sec:results}
\subsection{Creative Process}
\label{sec:results_process}

Overall, \textsc{LoopLens} supported broad exploration while enabling targeted refinement. Below we report results on 1) search behavior, 2) composition strategies and effort, 3) viewing/UI preferences that shaped how participants navigated the interface, and 4) participants' interviewed response of perceived task difficulty. 

\begin{figure}[htbp]
    \begin{center}
        \includegraphics[width=0.9\linewidth]{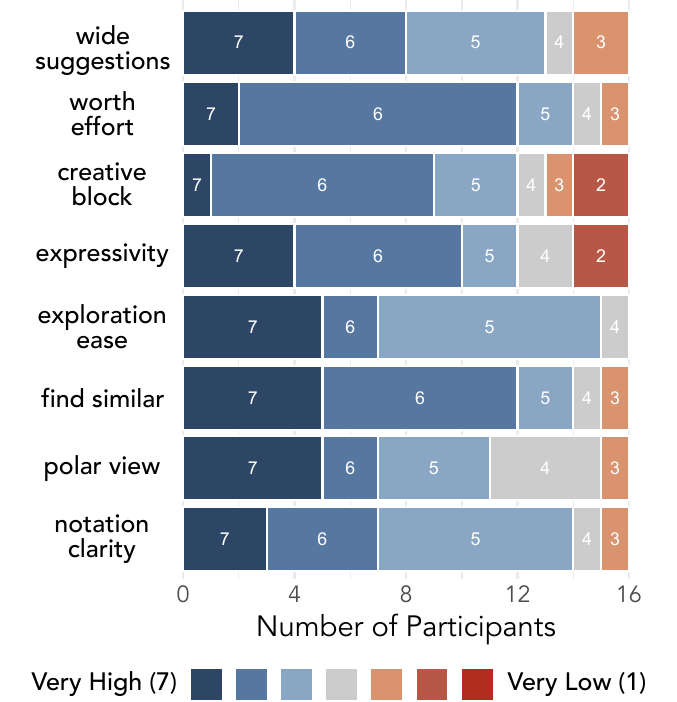}
    \end{center}
    \caption{Distribution of participants’ ratings on self-perceived experience using \textsc{LoopLens}.}
    \label{fig:post_survey}
\end{figure}

\subsubsection{Search Supports Broad Exploration.}
\label{sec:finding-search}

\begin{figure}[h!]
    \centering
    \begin{subfigure}[b]{0.45\textwidth}
        \centering
        \includegraphics[width=\textwidth]{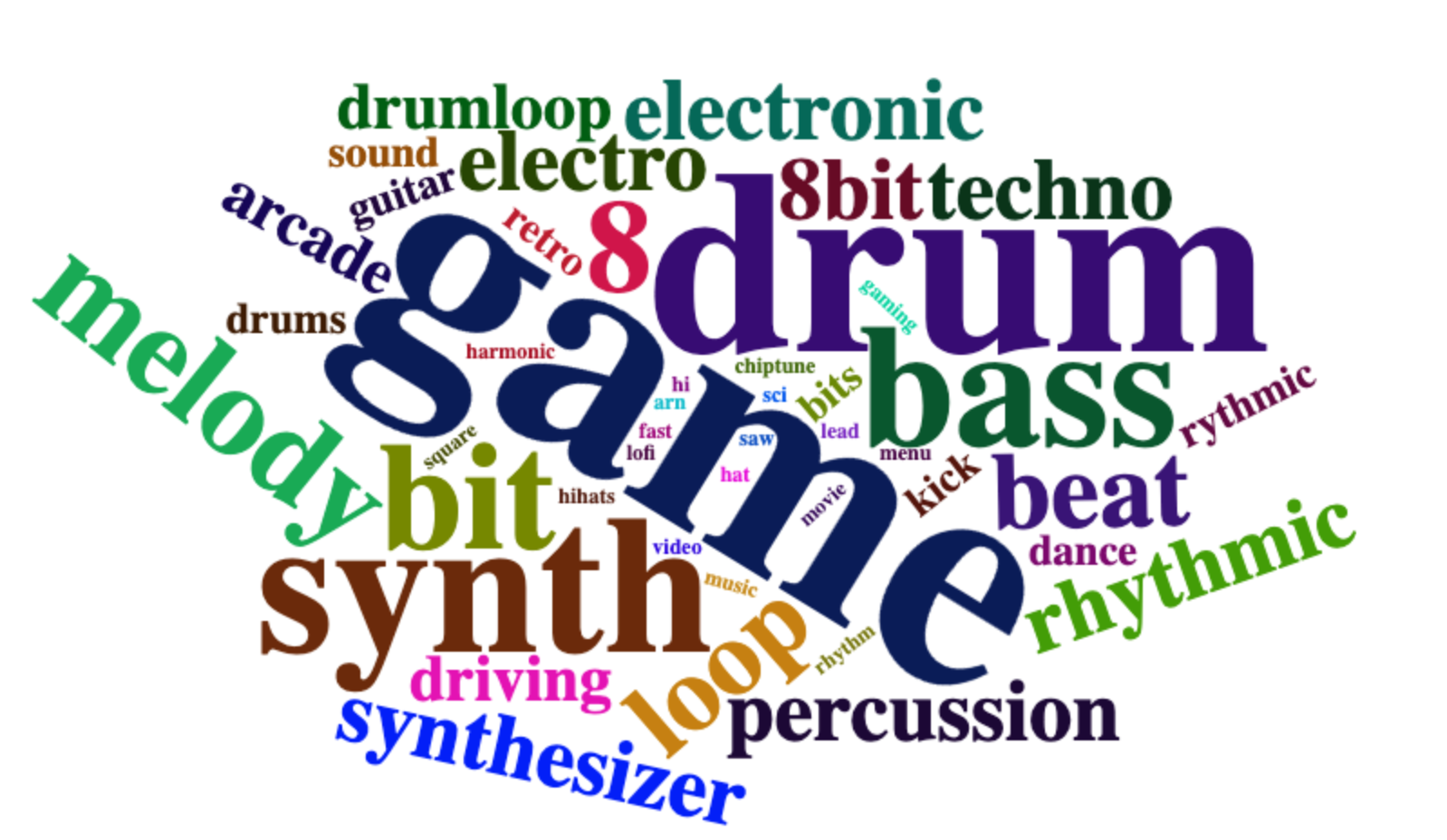}
        \caption{}
        \label{fig:sub1}
    \end{subfigure}
    \begin{subfigure}[b]{0.45\textwidth}
        \centering
        \includegraphics[width=\textwidth]{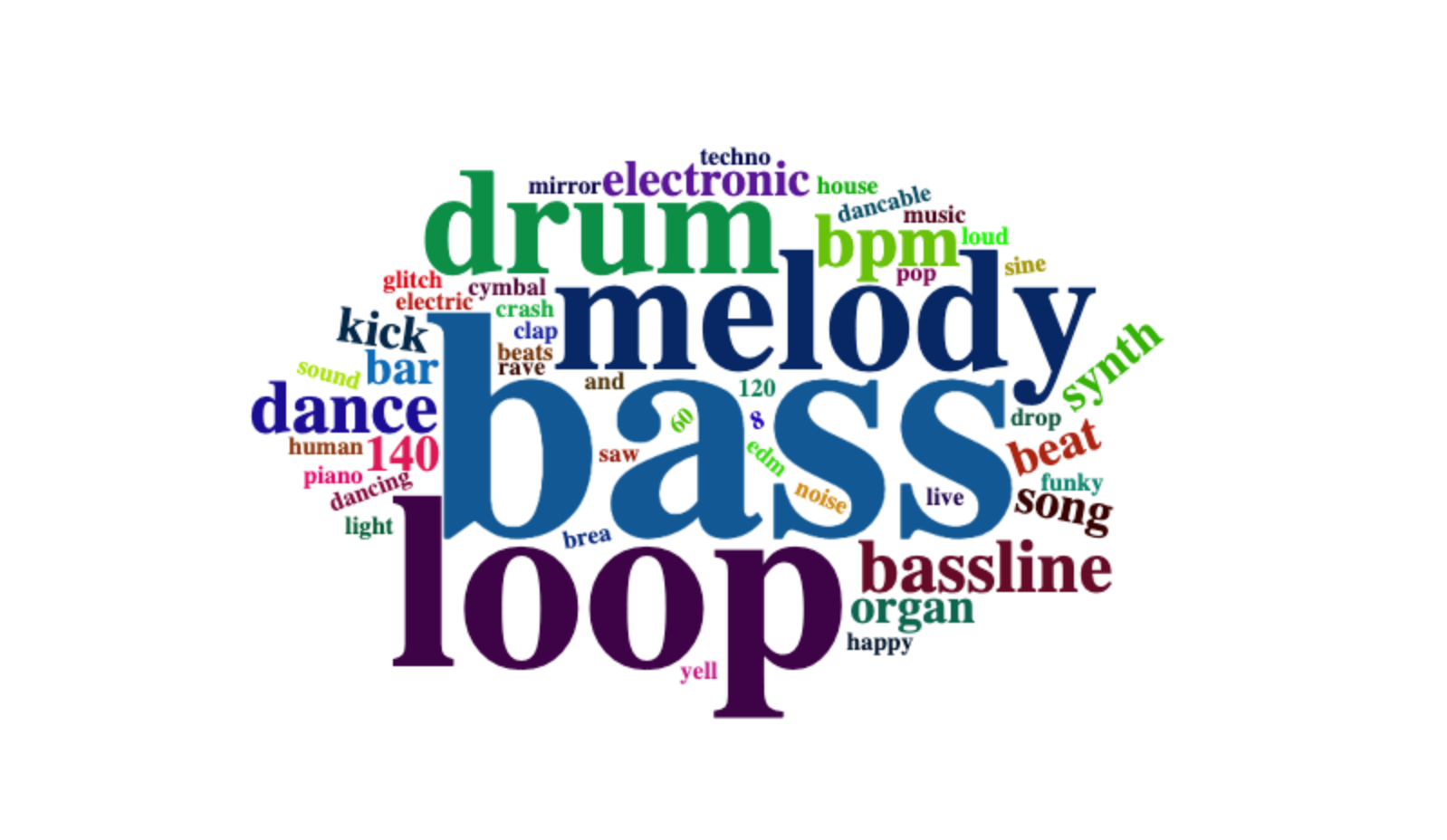}
        \caption{}
        \label{fig:sub2}
    \end{subfigure}
    \caption{Word clouds for both the (a) convergent task and the (b) divergent task. Notice that ``Chiptune'' appeared in search queries for the convergent task even though it did not appear in the tags provided on the reference song page.}
    \label{fig:wordcloud}
\end{figure}

Participants found \textsc{LoopLens} effective for surfacing options they would not have generated on their own (\texttt{wide suggestions} in \Cref{fig:post_survey}, $5.44 \pm 1.31$) and for making it easy to explore different ideas and outcomes (\texttt{exploration ease} in \Cref{fig:post_survey}, $5.69 \pm 1.01$). Quantitatively, participants submitted more queries for the convergent task (median = $7$, $IQR=6$) than the divergent task (median=5, $IQR=2$), a significant difference ($V=55.5, p=0.024$). For convergent searches, participants often drew directly from tags associated with the reference track, using terms like ``game'',  ``8bit'', and ``arcade'' (\Cref{fig:sub1}), while in divergent tasks they relied on broader terms such as ``loop'' to access the full database (\Cref{fig:sub2}), or ``bass'' that is often emphasized in modern music production of EDM. 


\textbf{\textit{Experts Use Filters as Constraint Satisfaction.}}  
Filtering was a distinctive marker of expertise. Experts applied significantly more filters per query (median=$0.97$, $IQR=0.86$) than novices (median=$0.10$, $IQR=0.26$; $W=29, p=0.0067$). Only experts used key filters, typically after committing to an initial track and searching for compatible additions. As \faGlasses{}~\texttt{P7} stated, this turned the task into a constraint satisfaction problem of findings keys that match each other: \textit{``... at a certain point, as soon as the first non-rhythmic stem was decided upon, it became like a constraint satisfaction problem.''}. 

\textbf{\textit{Diverse Usage of ``Find Similar''.}}  
The ``Find Similar'' feature received positive ratings (\texttt{find similar} in \Cref{fig:post_survey}, $5.88 \pm 1.15$) but usage patterns varied widely. \added{For novices, the feature acted as a ``warm start''. Participants who lacked the terminology to describe specific electronic sounds (e.g., ``chiptune'') relied on similarity recommendations to navigate the sample space without needing to formulate text queries on the spot. As \faBaby{}~\texttt{P14} noted, they could ``listen to everything'' related to a seed sound, effectively bypassing the keyword bottleneck.}

\added{For experts, the feature introduced an \textbf{interpretability} barrier. While some experts engaged with the feature out of curiosity regarding the CLAP embeddings, others struggled to map the algorithmic similarity score (e.g., ``92\%'') to their musical mental models. As \faGlasses{}~\texttt{P1} noted, ``I didn't know how to interpret the number that was [given]'', and \faGlasses{}~\texttt{P7} commented that ``it really matters what you mean by similarity''. This suggests a form of ``curse of knowledge'': experts possess a multi-dimensional understanding of similarity (encompassing harmony, rhythm, and timbre) and found the single scalar value, derived from audio embedding distance, opaque. Despite these different entry points, users who engaged extensively with the feature converged on a common request: \textbf{more control}. Several participants (\faGlasses{}~\texttt{P4}, \faBaby{}~\texttt{P6}) asked for additional filtering within similarity results, such as sorting the top-10 by BPM or key, to make recommendations more actionable.}\deleted{Some participants used it extensively ... they also wanted more transparency and control.}




\textbf{\textit{Deciding What to Insert Still Relied on the Ear.}}  
Ultimately, most participants judged suitability by listening. Metadata such as key and BPM played a \added{supporting}\deleted{secondary} role\added{, acting as a high-level filter}: some hovered for details before inserting, but few relied on metadata alone. As \faGlasses{}~\texttt{P7} explained: \textit{``I think the whole stuff was useful. It was just I could tell as soon as I started listening to this stuff, I was like, yeah, well, your best signal is your ears, you know?''} \added{However, the reliance on auditory intuition manifested differently across expertise levels. While experts used their ears to verify metadata constraints, novices used listening as their primary differentiation mechanism. \faBaby{}~\texttt{P14} stated: \textit{``I didn't really pay attention to the texts [of the titles] ... I'd search by tags and listen to everything I found [from that tag] in one sitting''.} }This underscores that auditory intuition remained central\added{ regardless of prior musical expertise}, with metadata serving mainly as a confirmatory or narrowing device\added{ for users who possessed the domain knowledge to interpret harmonic and rhythmic compatibility}.



\subsubsection{Composition: Rewarding, But Not Always Effortless}
\label{sec:finding-composition}

Participants reported that composing with \textsc{LoopLens} was generally worthwhile, rating their outputs as worth the effort (\texttt{worth effort} in \Cref{fig:post_survey}, $5.69 \pm 1.01$) and the activity as expressive and creative (\texttt{expressivity} in \Cref{fig:post_survey}, $5.38 \pm 1.63$). They only slightly agreed that they experienced creative block (\texttt{creativity block} in \Cref{fig:post_survey}, $5.06 \pm 1.53$), suggesting that while the process was not without challenges, it was rarely paralyzing. \deleted{As \texttt{P12} stated: \textit{``I feel like there’s a lot of things I want to incorporate into the song, but it either doesn’t match with the start or the overall scene of the divergent task.''}}

\textbf{\textit{Experts Hesitate Less, Perfectionists Stall.}}  
We measured the absolute time difference between the first preview to the first DAW insertion as a proxy for hesitation. Experts inserted\added{ their first candidate loop} significantly faster than novices (Median = $39.2$s vs $243.1$s; $W = 42, p = 0.01$). Outliers \faBaby{}~\texttt{P9} and \faGlasses{}~\texttt{P12} delayed over ten minutes before inserting, both ultimately submitting single-track composition for the convergent task.\added{ Qualitative evidence suggests this delay stemmed from imposing rigid internal evaluation criteria regarding what constituted a valid ``match'' for the prompt.} As \faBaby{}~\texttt{P9} noted: \textit{``I go for the perfect [match].''}\added{ Similarly, \faGlasses{}~\texttt{P12} rejected multiple options because they conflicted with their specific mental model: \textit{``I feel like there’s a lot of things I want to incorporate into the song, but it either doesn’t match with the start or the overall scene of the divergent task.''}} These cases illustrate how creative block manifested as prolonged hesitation rather than complete stoppage\added{, likely driven by an over-active evaluation process that inhibited initial commitment}.

\deleted{\textbf{\textit{Three Distinct Composition Strategies Emerged.}}  
Participants approached ...  approaches to balancing exploration and refinement.}


\textbf{\textit{Constraints Shaped (and Sometimes Blocked) Creativity.}}  
Despite resourceful strategies, participants often encountered limits in the dataset and tool. \faGlasses{}~\texttt{P7} remarked: \textit{``There’s not enough things to choose from, to be honest.''} \faGlasses{}~\texttt{P12} emphasized the lack of editing: \textit{``Since there is no editing, I cannot make little changes.''} Others observed that loops with odd bar counts were unusable or that transitions between segments felt awkward. These constraints sometimes led to circular searches, as when \texttt{P7} reopened the same similar-modal results and quipped: \textit{``This is where I was.''} Such issues help explain why participants reported mild creative block despite rating the overall process as rewarding. \faGlasses{}~\texttt{P1} stated that they felt at times stuck when they \textit{``... found a loop that sounded nice, but then I was unable to find another loop of the same key.''} or \textit{``I could not loop it because it could not fit 4 bars exactly''}. This same idea was echoed by \faGlasses{}~\texttt{P12}, on the lack of ability to edit\added{ or trim samples that span an odd number of bars: \textit{``There are a lot of tracks but usable ones are pretty rare, for example they cannot be 3 or 5 or 7 where they can't be evenly looped.''} Interestingly, however, we also observed workarounds from \faGlasses{}~\texttt{P4}, who encountered samples with trailing cut-offs but chose to layer them such that the downbeat of a new sample masked the tail of the previous one, creating a manual pseudo-crossfade effect. This behavior aligns with Gurevich et al.'s findings on how constraints can force stylistic variation and new technique generation~\cite{gurevich2012playing}.} 

\added{\textbf{Three Distinct Composition Strategies.} Synthesizing our behavioral logs and screen recordings, we observed three distinct composition strategies that illustrate different approaches to managing the exploration-exploitation tradeoff inherent in information foraging behaviors.} 

\added{Some participants (e.g., \faBaby{}~\texttt{P6}) extensively used the bookmarking feature to curate a palette of candidate sounds before ever touching the DAW timeline. This strategy decouples exploration from exploitation by introducing an intermediate \textbf{``soft commit''} --- sounds are marked as promising without being locked into the composition. This reduces cognitive load by narrowing the decision space incrementally, allowing users to separate the ``search'' mindset from the ``arrangement'' mindset.}

\added{A second group (e.g., \faGlasses{}~\texttt{P5} and \faGlasses{}~\texttt{P7}) engaged in \textbf{tighter exploration-exploitation cycles} and treated the search results and the timeline as a fluid, singular workspace. They frequently dragged loops directly from search to timeline, tested them in context, and immediately deleted or replaced them (\faGlasses{}~\texttt{P7}: ``Let me just try this for giggles''). Here, exploration and exploitation occur in rapid alternation rather than distinct phases, as each insertion is both a test and a tentative commitment.}

\added{The third strategy, predominantly observed in novices (e.g., \faBaby{}~\texttt{P16}), involved rapid-fire playback of search results without bookmarking or immediate insertion. These users engaged in \textbf{prolonged exploration}, relying on high-speed auditory scanning rather than metadata filtering to evaluate candidates. Two factors likely drive this pattern: lacking the musical vocabulary to efficiently constrain the search space, novices must evaluate more options serially; additionally, users with high internal thresholds for what constitutes a ``match'' may resist soft commitment until encountering a visceral fit.}

\added{These strategic differences highlight that users vary not only in \textit{what} they search for, but in \textit{how} they structure the search-to-composition transition. While some of these strategies are partially supported by current industry-standard DAWs, \Cref{sec:design-discovery} discusses design implications for supporting more exploration-driven workflows.}

\subsubsection{View \& UI}
\label{sec:finding-view-ui}

\begin{figure}[htbp]
  \begin{center}
    \includegraphics[width=0.95\linewidth]{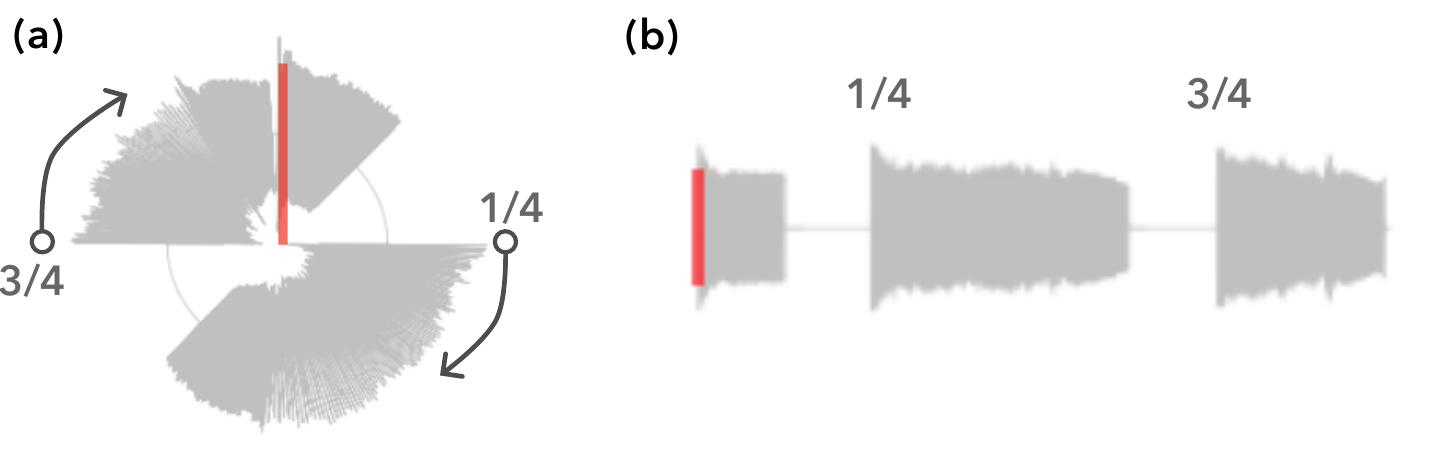}
  \end{center}
  \caption{\faGlasses{}~\texttt{P12} using \textit{Slider 140bpm} by Clinthammer (\url{https://freesound.org/s/179542/}) as an example. (a) and (b) are the radial and linear waveforms of the same song.}
  \Description{}
  \label{fig:P12-example}
\end{figure}

Overall, participants rated the radial waveform view as helpful for understanding the rhythmic qualities of the musical loops (\texttt{polar view} in \Cref{fig:post_survey}, $5.38\pm1.36$) and judged that where aspects of the notation mean similar things, the similarity is clear in the way they appear favorably (\texttt{notation clarity} in \Cref{fig:post_survey}, $5.44\pm1.09$).

\textbf{\textit{Polar Charts Are Helpful, But Only for Some.}}  
Preferences for polar audio waveforms split sharply across participants. Six of the 14 participants never used the polar view for either task, 4 of whom were experts. Among these, one expert relied exclusively on the list view, while the others stuck with the default combination of grid $+$ linear waveform views. Participants who did use polar plots generally found them beneficial. For instance, \faGlasses{}~\texttt{P4} described how they supported loop analysis: \textit{``having the polar plots to visualize how sparse the audio is or how busy it is, especially for separating the parts with chords and melodies to those that are just percussion was helpful.''} Similarly, \faGlasses{}~\texttt{P12} noted that \textit{``without listening to the soundtrack itself, we know what the sound would approximately look like, especially in a looped format''}, pointing to \Cref{fig:P12-example} as an illustration.\added{ Interestingly, \faBaby{}~\texttt{P13} offered a unique perspective --- despite lacking music theory training and prior exposure to radial charts, they found the circular waveforms intuitive due to their signal processing background. This suggests that high \textit{visualization literacy}~\cite{boy2014principled} allows users to successfully transfer visual metaphors~\cite{ziemkiewicz2008shaping} even when encountering novel layouts.}

Participants who avoided polar plots primarily cited \deleted{ or} familiarity with linear waveforms\added{ as the reason}. As \faBaby{}~\texttt{P15} explained: \textit{``I'm more used to reading linear plots, and because I haven't seen polar plots that much, I'm not used to it.''} Experts in particular emphasized that their prior experience with DAWs and waveform editing tools made the linear view more efficient. These patterns suggest that expertise and existing tool conventions create adoption barriers for novel representations, even when participants rate them positively.

Interestingly, polar charts sometimes introduced alternative visual metaphors~\cite{ziemkiewicz2008shaping}. For example, \faBaby{}~\texttt{P2}, who had no prior DAW experience, initially believed the polar view depicted the \emph{entire} sound segment while the linear view represented only a \textit{fraction}, despite both views spanning the same time duration\added{ for the same song}. This misinterpretation highlights how radial arrangements may cue metaphors of completeness or cyclicality that differ from linear representations.

\textbf{\textit{Grid View Is Preferred for Density.}}  
Across participants, the grid layout was favored over the list view, even though both displayed the same number of results (20 per page). Participants explained this preference in terms of information density: as \faGlasses{}~\texttt{P8} remarked, the grid view \textit{``... is more compact, so it contains more information.''} This suggests that layout choices shaped perceived richness of search results, with participants valuing compact displays even without changes to the underlying content.

In summary, while polar views were seen as conceptually helpful, their actual uptake was uneven, strongly moderated by prior expertise and familiarity. Grid layouts, in contrast, were broadly preferred, with participants gravitating toward denser result presentations regardless of content. 

\subsubsection{Perceived Task Difficulty \& Creativity}
\label{sec:finding-task-diff}

\added{Given the limited size of the database, we expected that participants would universally find the convergent task more difficult; yet, s}\deleted{S}omewhat unexpectedly, not all participants judged \added{it as such}\deleted{the convergent task as harder}. For \faBaby{}~\texttt{P16}, perceived difficulty depended on \textit{task order}: \textit{``If the task two would have been first time, it would have been also difficult to me.''} Among the remaining 15 participants, three (\faBaby{}~\texttt{P2}, \faGlasses{}~\texttt{P8}, \faBaby{}~\texttt{P15}) reported that the divergent task was more difficult. \faBaby{}~\texttt{P2} explained: \textit{``As a layman, the second test is easier 'cause I know I have something to follow.''} \faBaby{}~\texttt{P15} struggled to translate a single image into search terms: \textit{``One single image information ... it's too ambiguous for me to create the music. I do not know how to define the keywords. The only one I had in mind was `electronic,' nothing else.''} For \faGlasses{}~\texttt{P8}, difficulty came from knowing exactly the sound they wanted but finding the library lacked it. By contrast, the majority of the participants emphasized that the convergent task was harder because as \faBaby{}~\texttt{P13} states, \textit{``it's giving me a target. I am trying to minimize the error''}. Others noted the frustration of limited matches despite multiple queries due to the limited sound sample size. For example, \faGlasses{}~\texttt{P12} states that ``\textit{I think like if there's a larger datasets, maybe [the convergent task] can be easier.''}

Taken together, these perspectives suggest \added{another}\deleted{a} form of ``curse of knowledge.'' Novices found divergent tasks difficult because they lacked strategies for generating useful keywords, while more experienced participants found them difficult because they recognized when the database lacked the sounds they envisioned. Conversely, the convergent task posed challenges of precision and constraint, especially for perfectionist participants.

\subsection{Creative Outcome}
\label{sec:finding-creative-outcome}

Following the approach detailed in \Cref{sec:method_evaluation}, we conducted a homogenization analysis of participants’ final compositions by embedding each piece using pre-trained audio representation models and computing the cosine similarity of each embedding to the group mean embedding. 

\begin{figure*}[!htbp]
    \centering
    \includegraphics[width=0.75\linewidth]{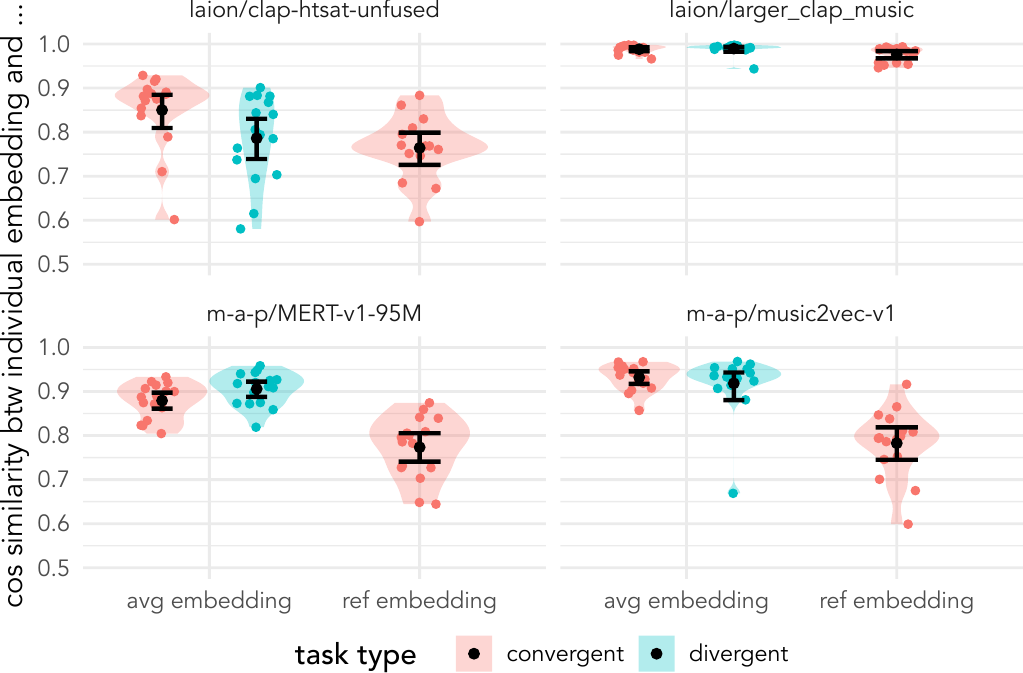}
    \caption{Homogenization analysis of participant's composition with bootstrap confidence interval overlaid on top. Error bars indicate resampled uncertainty around the mean.}
    \label{fig:output-total}
\end{figure*}

From \Cref{fig:output-total}, we observe that \texttt{to-avg-embedding} similarities were uniformly high across all models ($>0.74$ for convergent, $>0.80$ for divergent), with narrow bootstrapped confidence intervals. This right-skewed distribution across all four pre-trained weights likely reflects a homogenization effect: because our dataset is limited (331 loops) and stylistically coherent, embeddings are pulled toward a shared centroid, inflating average similarity. In contrast, \texttt{to-ref-embedding} comparisons revealed clearer separation across models (e.g., \texttt{clap-htsat-unfused} at $0.71 [0.64, 0.77]$ vs. \texttt{larger\_clap\_music} at $0.98 [0.97, 0.98]$), suggesting that reference-based evaluation is more sensitive to model differences.

Interestingly, the \texttt{larger\_clap\_music} model produced embeddings that were tightly clustered, reducing the ability to detect differences across participants. We hypothesize this is because the \texttt{larger\_clap\_music} is trained to capture broad music semantics, while EDM loops --- sharing meter, tempo ranges, and production conventions --- tend to be embedded very closely, yielding high cosine similarity even when stylistic nuances differ. 
This highlights that model choice strongly affects homogenization outcomes: while larger models may capture broad semantic similarity, they can obscure fine-grained creative variation. For future evaluations, models that preserve variance across stylistic or structural dimensions may be more informative. This will likely require within-genre contrastive training, beat-synchronous multi-scale representations, and auxiliary rhythm/timbre objectives that preserve fine-grained variation rather than smoothing it away.



\section{Discussion}

\subsection{\added{Design Implications}}

\added{Our findings offer implications for both creativity support tools and professional music production software (some of which are already emerging in recent software updates).} 

\subsubsection{\added{Task Design for CST Evaluation}}
\added{Our mixed findings regarding task difficulty highlight the complex relationship between task constraints and individual creative processes (\Cref{sec:finding-task-diff}), similarly observed in prior work on constrained instruments~\cite{gurevich2012playing}.
This suggests that CST evaluations relying solely on open-ended divergent tasks, common in prior work~\cite{louie2020novice,louie2022expressive}, may miss important aspects of how tools support goal-directed creative work. Future CST research should systematically vary task characteristics (constraint type, reference material modality, evaluation criteria) to understand how they interact with user factors (expertise, creative thinking style) in shaping both process and outcome.}
\subsubsection{\added{Automated Creativity Assessment}}
\added{Our multiverse analysis across embedding models revealed that creativity measurement is contingent on the underlying representations used (\Cref{sec:finding-creative-outcome}).
For CST researchers adopting embedding-based homogenization analysis~\cite{anderson2024homogenization}, this suggests careful attention to model selection: general-purpose audio models may smooth away the fine-grained variations that matter for assessing creativity within a genre or style. Domain-specific models, or models trained with objectives that preserve rhythmic and timbral variance, may prove more informative.}

\subsubsection{\added{Vocabulary Scaffolding for Novices, Proactive Compatibility Surfacing for Experts}}
\added{Novices' struggles with query formulation (\Cref{sec:finding-task-diff})
suggest value in vocabulary support mechanisms. This aligns with Andersen and Knees~\cite{andersen2016conversations}, who argue that effective semantic retrieval is ``not only a matter of just tags and words, but rather an ability to stay much closer to the vocabulary and mental representations of sound of each user''. These might include personalized semantic calibration during onboarding. By asking users to audition reference samples and place them on subjective scales (e.g., dragging a slider between ``bright'' and ``dark''), the system could learn a personalized mapping between the user’s internal descriptors and the database's audio features, effectively translating the user's unique mental model into retrievable queries. On the other hand, experts' use of filters as constraint satisfaction (\Cref{sec:finding-search})
points toward more proactive compatibility features. DAWs could highlight loops matching the key or tempo of existing tracks, or allow secondary filtering within similarity recommendations
to better support expert workflows where harmonic and rhythmic compatibility are non-negotiable.}


\subsubsection{\added{Preventing Cyclical Search in Similarity Navigation}}
\added{Similarity-based recommendation creates a structural risk: because similarity is symmetric, users exploring via ``Find Similar'' can inadvertently cycle through a small neighborhood of related sounds without realizing they are revisiting the same territory (\Cref{sec:finding-search}).
Without external memory of what has been auditioned, users may circle through sounds without finding an exit toward semantically meaningful novel material. Design interventions could include 
``already heard'' indicators on search results, or asymmetric recommendation strategies that progressively surface more distant sounds after repeated queries within a local neighborhood. 
}


\subsubsection{\added{From Retrieval to Discovery}}
\label{sec:design-discovery}
\added{Our framing of ``search as creation'' also aligns with an architectural shift emerging in tools like Ableton Live 12's Splice integration~\cite{abletonRelease}, moving from local retrieval to online discovery. 
Our observation of diverse composition strategies (\Cref{sec:finding-composition}) suggests this shift particularly benefits users engaged in extended exploration. Future tools could further support extended exploration through adaptive interfaces that recognize scanning behavior and adjust information density accordingly.
This shift --- from managing what you have to discovering what you might want --- treats the search interface not as infrastructure but as a creative space in its own right. While such integrations could introduce new visual metadata (e.g., cover art), our findings regarding the preference for the Grid View (\Cref{sec:finding-view-ui}) suggest that users will likely continue to prioritize high information density to facilitate rapid scanning.}




\subsection{Limitations \& \deleted{Potential }Future Work}

\subsubsection{Technical Limitations} 
\deleted{Our study faced several technical constraints that suggest directions for future work.} The current interface lacked granular editing capabilities (e.g., trimming, effects processing), which may have constrained participants' creative expression. Additionally, our interaction logging captured only the most recent search rather than comprehensive exploration histories, limiting our understanding of participants' exploration strategies. \deleted{Future iterations should implement more sophisticated provenance tracking to capture the full creative journey.}

\subsubsection{Methodological Limitations}
The proprietary nature of music data and sparse annotation quality limited our dataset options\deleted{, echoing broader challenges in computational creativity research}. Our 16-participant sample\added{ with participants recruited from a single company}, while sufficient for qualitative insights, constrains generalizability of \deleted{quantitative }findings. Future work would benefit from larger-scale studies\added{ with participants from more varying backgrounds} and improved \added{musical }datasets with richer semantic annotations.

\subsubsection{Measurement Validity}
Our creativity metrics, while grounded in established approaches, represent just one perspective on musical creativity. The reliance on embedding-based similarity assumes that distance from group averages captures meaningful creative divergence, an assumption that merits further validation through expert evaluation and alternative creativity measures.

\subsubsection{\added{Future Directions}} 
\added{Several directions emerge from this work: Our study captured single-session behavior; understanding how users develop search strategies and acquire domain vocabulary over time requires longitudinal investigation. Also, improving similarity transparency could increase adoption of features like ``Find Similar'' --- future interfaces might expose the dimensions along which similarity is computed, allowing users to weight audio similarity against metadata compatibility.}

\added{Last but not least, our findings raise questions about AI-mediated creative support that extend beyond similarity recommendation. The "Find Similar" feature represents a lightweight form of AI assistance, and participants' varied responses --- from enthusiastic exploration to uncertainty about interpretation --- suggest a tension between serendipity and control. Recent work by Kosmyna et al.~\cite{kosmyna2025your} found potential cognitive costs of sustained AI assistance in writing tasks, with users showing reduced performance over time compared to non-users. While their context differs from ours, the finding raises important questions for creative tool designers: might AI scaffolding help novices in the short term while inhibiting development of independent search vocabulary and aesthetic judgment? Future work should explore how different levels of AI assistance, from similarity-based recommendation to granular control~\cite{louie2020novice,louie2022expressive} to fully generative composition, affect both immediate creative outcomes and long-term skill development. Longitudinal studies tracking vocabulary acquisition, search strategy evolution, and creative confidence would help clarify whether AI scaffolding serves as training wheels that users eventually outgrow or as crutches that atrophy underlying capacities.}



\deleted{Our mixed findings regarding task difficulty ... in creative preferences and strengths rather than assuming uniform task experiences.}

\deleted{From a pedagogical perspective ... general divergent thinking, or their integration?}

\deleted{Future research should explore how task characteristics ... assessment design and pedagogical approaches in creative domains.}





\deleted{The challenge of measuring musical similarity ... particularly for novice users who may rely on AI as a learning scaffold.}

\deleted{Our findings suggest a nuanced approach is needed. ... very creative capacities they aim to support.}

\deleted{Future work should explore how different levels of AI assistance ... particularly for novice creators who may become dependent on AI scaffolding without developing underlying creative competencies.}

\section{Conclusion}

In this paper, we \added{examined}\deleted{
introduced} the concept of ``search as creation'' to reframe the role of search within creativity support tools, particularly in domains like loop-based music composition where discovery and collage are central to the artistic process. We argued that for many creative practices, search is not a preliminary retrieval task but a core medium of composition itself.

Through the design and evaluation of \textsc{LoopLens}, a \added{research probe}\deleted{novel tool} that \added{implements common DAW search features in a controlled environment for systematic study}\deleted{tightly integrates search and composition through new visualizations}, we investigated how creative goals and domain expertise shape user behavior. Our user study, which employed both a conventional open-ended (divergent) task and a novel goal-directed (convergent) task, revealed a distinct behavioral dichotomy. We found that domain experts leveraged multimodal cues to efficiently exploit a narrow set of sounds, treating composition as a constraint-satisfaction problem. In contrast, novices relied primarily on audio-driven exploration, a process often constrained by a limited vocabulary for formulating effective queries.

This research offers several contributions. By foregrounding search as a creative practice, we provide an empirical foundation and a new lens for designing CSTs in assemblage-based \added{creative }domains. Furthermore, our introduction of a convergent task provides a new methodological approach for evaluating music CSTs, revealing unique challenges and strategies not captured by divergent tasks alone. Ultimately, our findings yield \deleted{clear }design implications for future tools, underscoring the need to support vocabulary-independent discovery to bridge the gap between novice exploration and expert exploitation. By treating the search interface not merely as a retrieval mechanism but as a creative canvas, we can build more effective and inclusive tools that empower users of all skill levels to \deleted{compose, }discover,\added{ curate,} and create.

\begin{acks}
We thank members of the Sony Group Corporation (SGC), Sony Electronics Corporation (SEC), and Sony Research for their feedback, support, and volunteer work.  
\end{acks}

\bibliographystyle{ACM-Reference-Format}
\bibliography{00-sample-base}

\appendix
\section{Glossary of Music Terms}

Here we provide a glossary of music terms, note that most of the terms related to EDM come from Butler~\cite{butler2006unlocking}. 

\begin{itemize}[leftmargin=*]
    \item Beat: Otherwise known as ``pulse''. The beat is the basic, steady, recurring unit of time in music. 
    \item \texttt{MIDI:} Stands for ``Musical Instrument Digital Interface''. It is a standard protocol that software and hardware devices use to send information to one another, like note information and parameter controls. When you plug a keyboard into your laptop to play sounds in your DAW, it most likely works via MIDI.
    \item \texttt{Meter:} Meter describes the number of beats in a measure (also know as a ``bar'') and how the beats are normally divided. 
    In other words, meter is the organization of beats into recurring patterns of strong and weak accents, forming measures (or bars). It tells you how the beats are grouped. It is represented by the time signature. There are simple meters, compound meters, and complex meters. 
    \item \texttt{Patch:} A combination of settings, saved to a file, which can be loaded into a device (like a synth or effect). Similar to a preset.
    \item \texttt{Preset:} Similar to a patch, a preset usually comes with a synth and is a combination of settings that can be loaded to recall a certain synth sound.
    \item \texttt{Sample:} A segment of audio used as a sound in a track. Samples can be kicks, snares, drum loops, fx, melodic lines, parts of a whole other song, etc.
    \item \texttt{Sample Pack:} A downloadable folder of samples all grouped together. They usually contain many different samples of different kinds.
    \item \texttt{Stems:} The individual audio tracks rendered from a finished track, usually used for remix purposes.
    \item \texttt{Synthesizer (Synth):} An electronic instrument that creates sound by using oscillators and a series of processing.
    \item \texttt{Take:} A singular recording of audio. Usually, multiple takes will be done to achieve a good recording.
    \item \texttt{Tempo:} The speed of a piece of music. It tells you how fast the beats are happening. Usually measured in ``beats per minute'' (BPM). 
    \item \texttt{Timbre:} Also known as ``tone quality'' or ``tone color''. In simple terms, timbre is what makes a particular musical instrument or human voice have a different sound from another, even when they play or sing the same note.
    \item \texttt{Time Signature:} The meter is represented by the time signature. 
    \item \texttt{Note:} A note is a symbol in musical notation that represents a specific musical sound, indicating its pitch and duration. Such as quarter note (\musQuarter) and whole note (\musWhole).
\end{itemize}

\section{DAW Screenshots}
\label{sec:daw-full}

\begin{figure}[!htbp]
    \centering
    \includegraphics[width=0.95\linewidth]{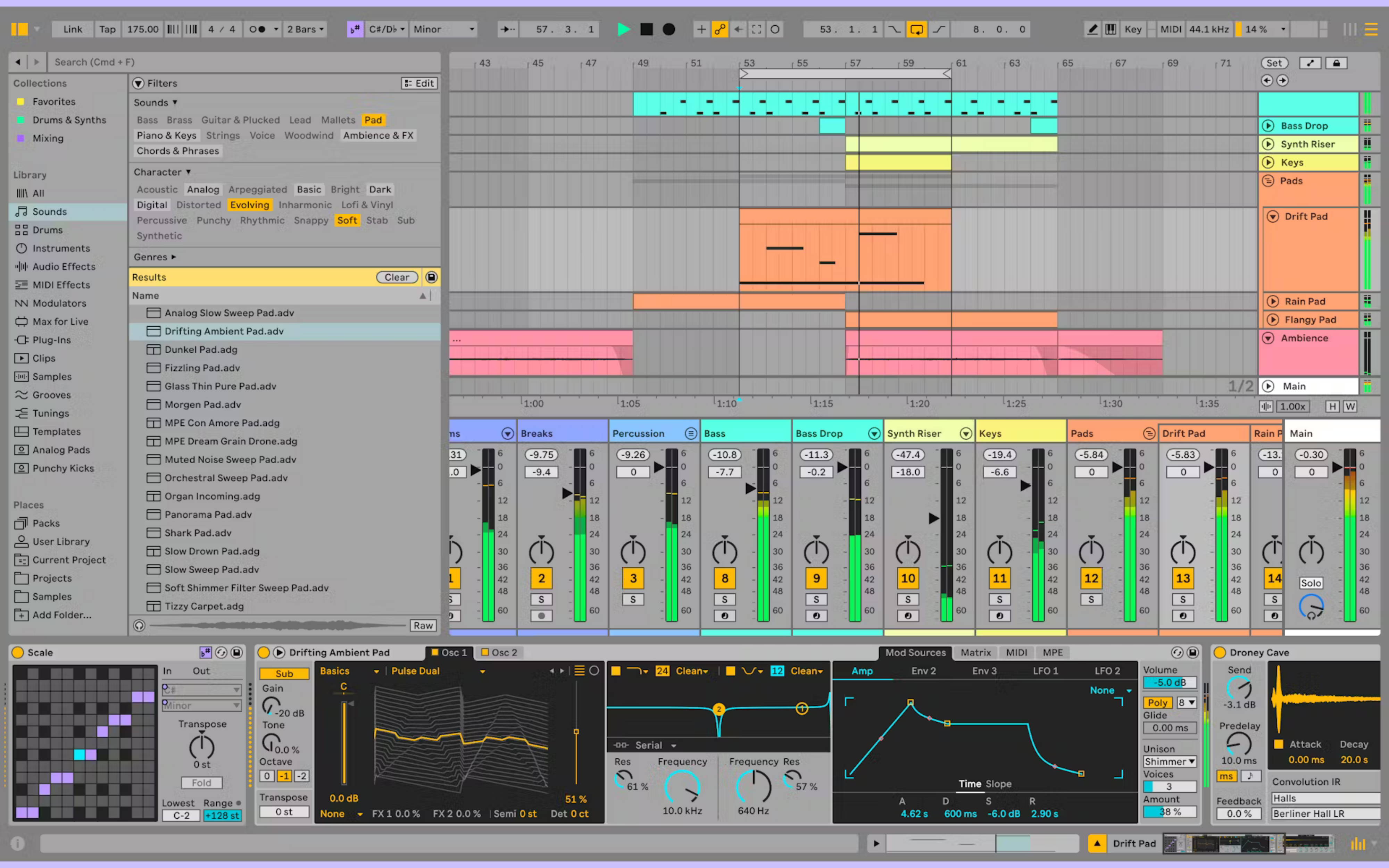}
    \caption{\added{Screenshot of Ableton Live 12's tool, taken from \url{https://www.ableton.com/en/live/}. Note that the browser is on the left.}}
    \label{fig:ableton-full}
\end{figure}

\begin{figure}[!htbp]
    \centering
    \includegraphics[width=0.95\linewidth]{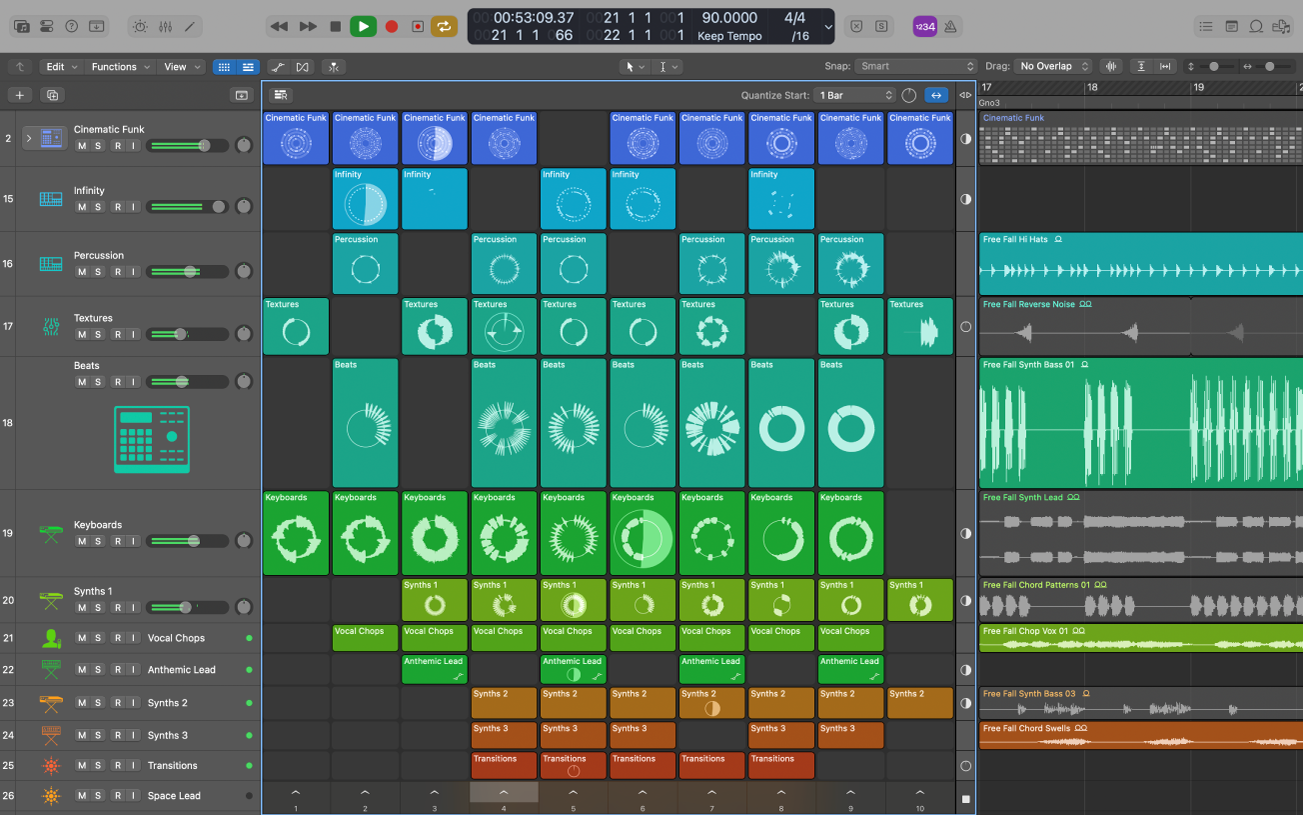}
    \caption{\added{Screenshot of Apple Logic Pro, taken from \url{https://support.apple.com/guide/logicpro/welcome/mac}. Note that the \texttt{Loop Browser} is currently collapsed but can be toggled via a button on the top left corner of the interface.}}
    \label{fig:logic-pro-full}
\end{figure}


\section{User Study}

\subsection{Survey Questions}
\label{sec:survey}

For the pre-task surveys in the user study, participants were asked the following questions: 
\begin{enumerate}[nosep,leftmargin=*]
    \item I have been practicing an instrument (or singing) on a daily and regular basis for the past \_\_\_ years. 
    \item At the time my interest was at its highest, I practiced my main instrument (including singing) for \_\_\_ hours per day. 
    \item I received \_\_\_ years of training in music theory. 
    \item I received \_\_\_ years of formal training in an instrument (or singing). 
    \item I can play \_\_\_ instruments. 
    \item The instrument I'm best at is \_\_\_. 
    \item My level of experience with \textbf{Digital Audio Work Station} (DAW) is \_\_\_. 
    \item Enter 10 words, each as different in meaning and association as possible. The more unrelated the words, the better. Here are some rules: 1) Only \textbf{single words} in English, 2) Only \textbf{nouns} (e.g., things, objects, concepts), 3) \textbf{No proper nouns} (e.g., no specific people or places), 4) \textbf{No specialized vocabulary} (e.g., no technical terms), 5) Think of the words \textbf{on your own} (e.g., do not just look at objects in your surroundings), 6) Try not to overthink it --- go with the first words that come to mind. 
\end{enumerate}

For the post-task surveys in the user study, participants were asked to rate their agreement with the following statements on a seven-point Likert scale (1 = Strongly Disagree, 7 = Strongly Agree).
\begin{enumerate}[nosep, leftmargin=*]
    \item Using \textsc{LoopLens}, I was able to receive a wide range of suggestions that I wouldn't have composed on my own. 
    \item What I was able to produce was worth the effort I had to exert to produce it. 
    \item I felt ``stuck'' or ``in a rut'' at some point during the creative process. 
    \item Using \textsc{LoopLens}, I was able to be very expressive and creative while doing the activity. 
    \item Using \textsc{LoopLens}, it was easy for me to explore many different options, ideas, designs, and outcomes. 
    \item The ``Find Similar'' feature provided a wide range of useful suggestions. 
    \item The radial waveform view helped me understand the rhythmic qualities of the musical loops. 
    \item Where aspects of the notation mean similar things, the similarity is clear in the way they appear. 
\end{enumerate}

\subsection{Interview Questions}
\label{sec:interview}

We list the questions used for the 15-minute semi-structured interview after the experiment session: 
\begin{itemize}[nosep, leftmargin=*]
    \item Can you give me your overall impression of using \textsc{LoopLens}? 
    \item What kind of \textit{search strategies} did you use for the first and second task? 
    \item What features did you find helpful and not helpful for the search strategies you employed? 
    \item Did you find the radial waveform view useful, and if so, in what way? If not, why? 
    \item Did you find the ``find similar'' button useful, and if so, in what way? If not, why? 
    \item Which of the two tasks felt more creative to you? Why? 
    \item Which task felt more difficult or challenging? In what way? 
\end{itemize}
\end{document}